\documentclass[11pt,letterpaper]{article}
\usepackage[T1]{fontenc}
\usepackage{lmodern,amsmath,amssymb,amsthm,mathtools,booktabs}
\usepackage[letterpaper,margin=1in]{geometry}
\usepackage[colorlinks=true,allcolors=black,bookmarksnumbered=true]{hyperref}
\newtheorem{theorem}{Theorem}[section]
\newtheorem{lemma}[theorem]{Lemma}
\newtheorem{corollary}[theorem]{Corollary}

\theoremstyle{remark}
\newtheorem{remark}[theorem]{Remark}
\newcommand{\bits}{\{0,1\}}
\newcommand{\F}{\mathbb F}

\newcommand{\Adr}{\operatorname{Adr}}

\DeclareMathSizes{10}{10}{7}{6}

\title{Sharp Bounds for Selector Advice and Oblivious Equilibrium Search}
\author{Sebastian Ben Daniel\\[2pt]
\small Ben-Gurion University of the Negev}
\date{November 2026}
\hypersetup{pdftitle={Sharp Bounds for Selector Advice and Oblivious Equilibrium Search},pdfauthor={Sebastian Ben Daniel},pdfsubject={Computational complexity}}
\begin{document}
\maketitle
\begin{abstract}
Ko's tournament argument recognizes a P-selective language with quadratic
advice. We determine the leading coefficient in the selector-query model.
On $n$-bit inputs, $n^2/4+O(n)$ ordinary advice bits and $n$ nonadaptive
comparisons suffice. Some cuts require $n^2/4-O(n\log n)$ bits even for
randomized decoders with adaptive access to individual-degree-two
extensions. At degree one the exact fixed advice width is $n+1$.
Assuming exponential hardness in $\mathrm E$ against SAT-gate circuits,
the quarter bound holds in deterministic polynomial time for every
P-selective language.

The same fixed point determines the cost of searching for an approximate
equilibrium before seeing the payoffs. In an $N$-action symmetric
constant-sum game with payoffs in $[0,1]$, let $P$ be the best worst-case
probability that one game-independent proposal is an additive
$\epsilon$-Nash equilibrium. For each fixed $0<\epsilon<1/2$,
\[
 -\log_2 P=(1+o(1))\,
 \frac{(\log_2 N)^2}{4[1-H_2(1/2+\epsilon)]}.
\]
The algorithm draws uniform actions and proposes their empirical
distribution for both players. The lower bound allows arbitrary mixed
profiles and $N^{o(1)}$ adaptive payoff inspections before the proposal.
For all symmetric two-player games, including general-sum games, the
sharp exponent is $(\ln 2/8+o(1))(\log_2 N)^2/\epsilon^2$ when
$\epsilon\to0$ and $\log(1/\epsilon)=o(\log N)$. For security strategies
in arbitrary zero-sum games the exponent is four times larger, and
sampling an optimal strategy attains it.

The common mechanism counts successful tuples. A fixed point makes them
capture constant sampling probability while controlling a power moment of their distribution.
\end{abstract}

\section{Introduction}
A selector is an algorithm that chooses one of two strings and chooses a member
of a language $L$ whenever either string belongs to $L$. Suppose the algorithm
runs in polynomial time. How much advice, common to all $n$-bit inputs, suffices
to decide membership in $L$? Ko's argument gives a short proof of a quadratic
upper bound~\cite{derand:Ko}. View the $n$-bit strings as the vertices of a tournament,
and direct each edge from the chosen string to the other one. Every member beats
every nonmember. Among any remaining members, some vertex is beaten by at least
half of them when equality counts. Store that vertex and repeat on the members
it fails to certify. The resulting list of at most $n+1$ names recognizes $L$
at length $n$. Its cost comes from writing down $n$ bits per name.

\paragraph{Prior work.}
Selman introduced P-selective sets as the polynomial-time analogue of the
semirecursive sets~\cite{Sel79}. Ko's quadratic bound has remained the
deterministic upper bound. Nondeterministic interpreters do much better:
Hemaspaandra and Torenvliet showed that $n+1$ advice bits suffice for
NP interpreters and that $n$ bits do not~\cite{HT96}. A linear deterministic
bound therefore follows from $\mathrm P=\mathrm{NP}$, and Hemaspaandra,
Hempel and Nickelsen derived it from the assumption that every P-selective
language has an associative selector~\cite{HHN04}. Domination numbers of
tournaments save only a lower-order term: the bound of Szekeres and
Szekeres~\cite{SS65,derand:LWW00} gives $n^2-n\log_2n+O(n)$ bits. In the
comparison model, our earlier work~\cite[Theorems~5.5 and 5.6]{BD} gave a lower
bound with leading coefficient $1/4$ for polynomially many adaptive Boolean
queries and an unrestricted-time upper bound with coefficient $1/2$.

We close this interval, extend the lower bound to algebraic queries of
individual degree two, and bring the upper bound to polynomial time under a
standard derandomization hypothesis. The hypothesis cannot be used to go
further by any relativizing argument.

\paragraph{The model.}
Put $N=2^n$. A legal cut consists of a tournament $E$ on $\bits^n$ and a
marked set $P$ all of whose vertices beat every vertex outside $P$. Write
$x\succ y$ for $E(x,y)=1$ and set $E(x,x)=0$. An asymmetric selector induces
this table by sorting its two arguments before evaluating it.
A decoder receives an input vertex, advice, and access to the table.
Its program is independent of the marked tournament. An algebraic query
specifies a finite field with its canonical encoding and a point of
$\F^{2n}$. The answer is the value of one fixed polynomial per field agreeing
with $E$ on the Boolean cube. Degree bounds are per variable. All queries
are classical. Between queries we allow unrestricted computation except in
results explicitly specifying polynomial time.

\begin{theorem}[The ordinary advice threshold]\label{main:ordinary}
Every legal cut has a zero-error decoder using $n^2/4+O(n)$ ordinary advice
bits and at most $n$ nonadaptive Boolean comparisons. One computable decoder
works for all lengths and cuts. Conversely, for each fixed randomized interpreter
making polynomially many adaptive queries and every sufficiently large $n$,
some cut and some fixed degree-two extensions require
$n^2/4-O_q(n\log n)$ ordinary advice bits for pointwise error $1/3$.
\end{theorem}

Here $O_q(\cdot)$ allows the constant to depend on the polynomial query bound.
The advice in the lower bound may depend on the entire realized oracle,
including its extensions, and the decoder's computation between queries is
unrestricted. The lower bound is therefore information-theoretic, and the
upper bound matches it with Boolean comparisons alone.

\begin{theorem}[Degree one]\label{main:degreeone}
With direct access to the canonical multilinear extensions, deterministic
polynomial time and $n+1$ fixed-width advice bits suffice for exact
recognition. This width is optimal even with the entire table available.
\end{theorem}

The decoder reads $\deg^+(x)$ from row evaluations: $x\in P$ precisely when
$\deg^+(x)\ge N-|P|$. One public odd prime greater than $N$ suffices for
a single query; a uniform algorithm uses $O(n/\log(n+1))$ small-prime queries.
Thus multilinear access gives a deterministic decoder exactly the advice
that NP interpreters need~\cite{HT96}, while access of individual degree two
already costs the full quarter. In the language of Aaronson and
Wigderson~\cite{AW08}, the advice complexity of selectors is a natural
quantity on which algebraic query access of degree one and of degree two
separate quadratically. Theorems~\ref{main:ordinary} and~\ref{main:degreeone}
are proved in Sections~\ref{fc:covers}, \ref{sec:alg-lower} and~\ref{fc:degree}.

\paragraph{Oblivious equilibrium search.}
Suppose two players have $N$ actions each. Before seeing the payoffs, we
choose a distribution on pairs of mixed strategies. What is the greatest
probability with which one draw is an approximate Nash equilibrium, even
for the worst game? The question measures the information needed to name
a successful strategy. With access to the full payoff matrix an explicit
zero-sum game can be solved by linear programming.

Daskalakis and Papadimitriou~\cite{DP09} called such procedures
oblivious: the algorithm samples a fixed distribution on pairs of mixed
strategies, and the game is used only to check whether the sample is an
approximate equilibrium. Sampling arguments of Alth\"ofer~\cite{Alt94}
and Lipton and Young~\cite{LY94} for zero-sum games, and of Lipton,
Markakis and Mehta~\cite{LMM03} for bimatrix games, give empirical
strategies of length $O(\log N/\epsilon^2)$. A uniform draw from the
resulting list of profiles succeeds with probability at least
$2^{-O(n^2/\epsilon^2)}$, where $n=\log_2N$. Babichenko, Barman and
Peretz~\cite{BBP14} studied simple approximate equilibria in games with
many players.

Daskalakis and Papadimitriou~\cite[extended version, proof of Theorem~4,
Claim~3]{DP09} show that finding a $\delta$-well-supported equilibrium
can require $\Omega(2^{(0.8-34\delta)n^2})$ oblivious proposals in
expectation for bimatrix games with payoffs in $[-1,1]$. Setting the
negative payoffs in their construction to zero gives the same coefficient
$0.8-34\epsilon$ for ordinary additive regret on our $[0,1]$ scale
(Remark~\ref{g:dp09-packing}). Kothari and Mehta~\cite{KM18} proved a
$2^{\Omega(n^2)}$ lower bound for ordinary regret at a sufficiently small
fixed error, even with access to a sum-of-squares relaxation and a partial
verification oracle. For symmetric constant-sum
games we determine the sharp coefficient for every fixed
$0<\epsilon<1/2$. It grows as $(\ln 2)/(8\epsilon^2)$ as $\epsilon\to0$.

\begin{theorem}[Equilibrium search]\label{main:game}
Let $P_\epsilon(N)$ be the optimal worst-case success probability of one
oblivious proposal for symmetric constant-sum games, normalized so
$A_{ij}+A_{ji}=1$ and $A_{ij}\in[0,1]$. Put $n=\log_2N$ and
$c_\epsilon=1-H_2(1/2+\epsilon)$. For fixed $0<\epsilon<1/2$,
\[
       -\log_2 P_\epsilon(N)=\frac{n^2}{4c_\epsilon}+O_\epsilon(n\log n).
\]
The same leading exponent holds for proposals with $N^{o(1)}$ adaptive
payoff queries, and for the logarithm of the smallest list that works
for every game. If $\epsilon\to0$ and $\log(1/\epsilon)=o(n)$, the
sharp exponent for oblivious proposals in the class of \emph{all}
symmetric bimatrix games is $(\ln 2/8+o(1))n^2/\epsilon^2$.
\end{theorem}

The factor four has a direct comparison. A column strategy is
$\epsilon$-secure in a value-$1/2$ game if every pure row earns at most
$1/2+\epsilon$ against it. For this same task, with the value known,
the logarithmic cost of oblivious search has leading term $n^2/(4c_\epsilon)$
in symmetric constant-sum games and $n^2/c_\epsilon$ in arbitrary
zero-sum payoff matrices (Theorem~\ref{g:general-sec}). The second
exponent is what sampling an optimal strategy achieves, as in the
arguments of Alth\"ofer and of Lipton and Young, so their sampling is
exactly optimal for oblivious search in zero-sum games. Symmetry saves
a factor four, the same factor that separates Ko's bound from
Theorem~\ref{main:ordinary}.

The proof chooses a distribution on actions and a sample length $r$
together. With this distribution, at least half the samples give a
secure strategy, and at least $2^{c_\epsilon r^2-O(r)}$ different
ordered samples succeed. This offsets $c_\epsilon r^2$ of the $nr$
bits needed to specify one sample. The worst length is near
$n/(2c_\epsilon)$. Conversely, a tournament with a hidden set of
about $\sqrt N$ winning actions makes this cost necessary even for
proposals using arbitrary real probabilities. Section~\ref{sec:games}
gives the sampler, and Appendix~\ref{app:games} proves the lower bound
and the extensions.

\paragraph{Bounded error and the cost of confidence.}
An ordinary advice word is fixed before the random tape is drawn, so
repeating a bounded-error decoder on fresh coins reuses the same word.
This is why the lower bound in Theorem~\ref{main:ordinary} already holds
at error $1/3$. Advice that may depend on the realized public tape behaves
differently: at constant error it permits linear advice~\cite{BD12}. How
the cost of such advice, and of coherent quantum advice, grows as the
error decreases is the subject of a separate paper. Nothing here depends
on it.

\paragraph{Polynomial time.}
Let $\mathrm H$ denote the hypothesis that for some $c>0$ a Boolean function
in $\mathrm E=\mathrm{DTIME}(2^{O(n)})$ requires circuits of size $2^{cn}$
with SAT gates on all sufficiently large input lengths. Our efficient
decoder uses the logarithmic-seed generator for SAT-gate tests supplied by
this hypothesis~\cite{derand:KvM,derand:Zim}.

\begin{theorem}[Efficient recognition under $\mathrm H$]\label{main:efficient}
Under $\mathrm H$, every P-selective language $L$ has a deterministic
polynomial-time decoder using $n^2/4+O_L(n)$ ordinary advice bits.
If the smaller nonempty side at length $n$ has $K$ vertices, the decoder
makes at most $\lceil\log_2 K\rceil+1\le n$ nonadaptive selector comparisons.
\end{theorem}

The computable and polynomial-time decoders thus attain the same additive
order above $n^2/4$. The constant hidden in $O_L(n)$ may depend on the fixed
selector. Earlier conditional improvements of Ko's bound assume
$\mathrm P=\mathrm{NP}$ or associative selectors. Theorem~\ref{main:efficient}
needs only a circuit lower bound of the kind used to derandomize
Arthur--Merlin protocols. The next theorem shows that this is the end of the
line for arguments that treat the selector as an oracle.

\begin{theorem}[Relativized sharpness]\label{main:oracle}
There are an oracle $O$ and a degree-two extension family $\widetilde O$ of
it such that $\mathrm H^O$ holds, every $\mathrm P^O$-selective language $L'$
has a polynomial-time decoder relative to $O$ with $n^2/4+O_{L'}(n)$ advice
bits, and some $\mathrm P^O$-selective language is not in
$\mathrm{BPP}^{\widetilde O}/\lfloor cn^2\rfloor$ for any $c<1/4$.
\end{theorem}

Consequently no relativizing proof can derive advice $(1/4-\delta)n^2$ for
P-selective languages from $\mathrm H$, and neither can a proof that
algebrizes in the sense of Aaronson and Wigderson~\cite{AW08} with
extensions of individual degree two. Section~\ref{fc:random-world-section}
gives the construction; it is Theorem~\ref{fc:random-world}.

\paragraph{What determines the cost.}
The strong components of a tournament are totally ordered, and a legal cut
falls between two consecutive ones. Only the smaller of these two
components matters.

\begin{theorem}[Boundary components]\label{main:component}
For a nontrivial legal cut let $B$ be the size of the smaller of the two
strong components adjacent to it, $v=\log_2B$ and $u=\min\{v,n/2\}$.
A computable decoder uses $u(n-u)+O(n)$ advice bits. For every
polynomial-query randomized decoder with degree-two access, every large
$n$ and every $3\le B\le2^{n-1}$, some cut with this boundary size requires
$u(n-u)-O_q(n\log n)$ bits at error $1/3$. For a P-selective language,
a deterministic polynomial-time decoder needs $n(\lfloor v\rfloor+1)+O(1)$
bits unconditionally and $u(n-u)+O_L(n)$ bits under $\mathrm H$.
\end{theorem}

When $\log n\ll v=o(n)$ the cost is therefore $(1+o(1))nv$, and it is
attained in polynomial time with no hypothesis. The quarter coefficient
already appears at $B=2^{\lceil n/2\rceil}$, and larger boundary components
do not raise it. The same exponent governs tournaments without any
reference to advice (Section~\ref{sec:tournaments}). In the worst case a
random $t$-subset of an $N$-vertex tournament dominates with probability
$2^{-(1/4+o(1))\log^2(N/t)}$, and every tournament on $\bits^n$ has a
dominating set of Kolmogorov complexity $n^2/4+O(n)$, which is tight.
The lower bound embeds the hard instances of
Theorem~\ref{main:ordinary} under a public block of padding vertices joined
to them by two private edges (Section~\ref{focus:component}).

\subsection{Where the quarter comes from}
Orient the smaller side $S$ so that it beats its complement. A list of vertices
of $S$ covers $S$ if every $x\in S$ equals or beats an entry. A uniform
ambient $r$-tuple has $nr$ bits in its description. If at least
$2^{r^2-O(r)}$ different tuples cover $S$, its chance of being a cover is at
least $2^{-r(n-r)-O(r)}$. A common list of candidate tuples can then serve
all tournaments, and it costs only $r(n-r)+O(n)$ bits to name a cover
in that list. The maximum of $r(n-r)$ is $n^2/4$. The central point is to find
a length $r$ at which both coverage and multiplicity hold.

For example, if a vertex loses to every other vertex of $S$, every cover
must contain it. Sampling uniformly from $S$ can take order $|S|$ draws
just to hit this one vertex. We give a vertex more weight when the current
distribution of anchors is likely to miss it. If $s_\nu(x)$ is the miss
probability, the fixed point has weights $p(x)\propto s_\nu(x)^J$ and uses
a slight smoothing $\nu$ of $p$. The normalization chooses $J$ so that
$r=\lceil J\rceil$ independent samples cover with constant probability.
The tournament identity also controls
$\sum_x\nu(x)^{1+1/J}$. H\"older's inequality then turns probability mass
on covers into the required count. Section~\ref{fc:covers} gives the proof.

The same count applies to any subtournament whose covers recognize the
cut, in particular to a boundary component, which gives the upper bound in
Theorem~\ref{main:component}.

To recover a cover in polynomial time, testing whether an entire tuple covers
$S$ would require certifying a statement about every vertex. Instead the
decoder's existence proof certifies how much of the remaining weight a
short block of anchors covers. This is a lower bound on the size of a
nondeterministically recognized set. One potential pays for uncovered
weight and another for the total description length. A lower cutoff on
vertex weights sets the block size: larger blocks save seeds, and the
cutoff keeps the necessary counting precision inverse polynomial.
The construction uses $O(n/\log n)$ blocks. Their seed and length fields
therefore cost $O_L(n)$ bits altogether. The decoder evaluates only the
resulting short descriptions (Section~\ref{fc:efficient}).

\subsection{Why the lower bound survives adaptive queries}
Partition the vertices into $2^m$ buckets. Each contains a hidden vertex
with a fair membership bit. Finding such a vertex in another bucket means
locating its $(n-m)$-bit suffix. An input's label can remain ambiguous
until it has been compared to several other hidden vertices. For ordinary
advice, amplification reuses the word and forces the cost up to
$m(n-m)-O_q(n\log n)$ bits. The balanced choice $m=\lfloor n/2\rfloor$
recovers the quarter.

An off-cube query may involve every entry of the table. We identify the
Boolean positions at which changing one entry can affect the likelihood
of its transcript. A degree-two polynomial that vanishes at all query
points can mask the change at every other position. Factoring such a
polynomial across the two endpoints of an edge shows that the transcript
of $q$ queries can address at most $2q$ vertices. There is also a stronger
conclusion across successive runs on the same oracle. If the earlier runs
never addressed the current input, any newly visible edge from it must have
its other endpoint addressed by the current run alone. This is the step
that permits an error count across inputs despite the shared extension.

\paragraph{Open problems.}
The unconditional lower bounds concern access to a selector table and
relativized languages. Without hypotheses, the ordinary advice complexity
of P-selective languages with polynomial-time decoding remains between
$n+1$ bits, forced by~\cite{HT96}, and $n^2-n\log_2n+O(n)$.
Theorem~\ref{main:oracle} shows that closing the gap below $n^2/4$, even
under $\mathrm H$, requires a nonrelativizing argument. It is also open
whether $\mathrm H$ can be replaced by a weaker derandomization hypothesis,
for instance one that only fools nondeterministic circuits of a fixed
polynomial size.

\section{Counting covering tuples}\label{fc:covers}
Orient the side to be recognized so that it beats its complement,
reversing all comparisons if needed. An anchor $a$ on this side
certifies $x$ when $x=a$ or $x\succ a$. A list covers the side when
it certifies each of its vertices. Checking a cover takes one
nonadaptive comparison per anchor. All logarithms are base two except
where $\ln$ is written.

We first give the fixed-point argument in a form that also applies to
payoff deviations in games. The exponent is a variable in the fixed
point. This allows the scores to exceed one before normalization.

\begin{lemma}[Normalizing a family of scores]\label{fp:scores}
Let $p\mapsto (g_i(p))_{i=1}^N$ be continuous on the probability simplex.
Suppose that the scores are positive on a common nonempty set of indices,
zero elsewhere, and that $\sum_i p_i g_i(p)\le 2^{-c}$ for every $p$, for
some $c>0$. Given $0<\zeta<1$, there are a distribution $p$ and a real
$0<J\le\log_2(N/\zeta)/c$ satisfying
\[
 \sum_i g_i(p)^J=\zeta,\qquad
 p_i=g_i(p)^J/\zeta,\qquad
 \sum_i p_i^{1+1/J}\le\zeta^{-1/J}2^{-c}.
\]
In particular every score at this distribution is less than one.
\end{lemma}
\begin{proof}
Restrict to the face supported on the positive indices, whose number is
$M\ge1$. Put $Z(p,J)=\sum_i g_i(p)^J$ on this face, summing over those
indices. Choose $U>\log_2(N/\zeta)/c$. On the product of the face with
$[0,U]$ apply Brouwer's theorem to the continuous map
\[
 (p,J)\longmapsto
 \left((g_i(p)^J/Z(p,J))_i,\
    \operatorname{proj}_{[0,U]}\bigl(J+\log_2(Z(p,J)/\zeta)\bigr)\right).
\]
At $J=0$ the sum is $M>\zeta$, so a fixed point cannot lie at this
endpoint. Whenever $J>0$ and the first coordinate is fixed, convexity gives
\[
 N^{-1/J}\le \sum_i p_i^{1+1/J}
       =Z(p,J)^{-1/J}\sum_i p_i g_i(p)
       \le Z(p,J)^{-1/J}2^{-c}.
\]
This forces $Z(p,U)<\zeta$ at a fixed first coordinate and excludes
the upper endpoint. The second coordinate at an interior fixed point
forces $Z=\zeta$. The same inequality proves both remaining bounds.
\end{proof}

\begin{lemma}[A self-consistent distribution of anchors]\label{fc:fixedpoint}
Let $S$ be a tournament on $K\ge2$ vertices, with uniform distribution $U$.
Fix $0<\epsilon\le1/2$ and $0<\zeta<1$. There are a distribution $p$
on $S$ and a real $J>0$ such that, for $\nu=(1-\epsilon)p+\epsilon U$
and $s(x)=\Pr_{a\sim\nu}[a\succ x]$,
\[
 \sum_xs(x)^J=\zeta,\qquad p(x)=s(x)^J/\zeta.
\]
Their power moment $A=\sum_x\nu(x)^{1+1/J}$ satisfies
\[
 A\le \tfrac12\zeta^{-1/J}(1+2\epsilon),\qquad
 J\le\frac{\log(K/\zeta)}{\log(2/(1+\epsilon))}.
\]
Furthermore $(\epsilon/4)^J\le4\zeta/K$.
\end{lemma}
\begin{proof}
Given $u$, let $g_x(u)$ be the miss probability under
$(1-\epsilon)u+\epsilon U$. Its positive coordinates are exactly the
vertices of positive indegree, regardless of $u$. For independent
$x,a\sim u$, the probability that $a\succ x$ is
$(1-\sum_xu(x)^2)/2$. Consequently $\sum_xu(x)g_x(u)\le(1+\epsilon)/2$.
Apply Lemma~\ref{fp:scores} with $2^{-c}=(1+\epsilon)/2$. It gives the
stated normalization and the bound on $J$. Convexity gives
\[
 A\le(1-\epsilon)\sum_xp(x)^{1+1/J}+\epsilon K^{-1/J}
 \le(1-\epsilon)\zeta^{-1/J}(1+\epsilon)/2+\epsilon K^{-1/J}
 \le \tfrac12\zeta^{-1/J}(1+2\epsilon).
\]
Finally, if $b>0$ vertices have indegree less than $K/4$, their mean
indegree in the induced tournament is $(b-1)/2<K/4$. At least $K/4$
vertices therefore have indegree at least $K/4$ (also directly for $K=2,3$).
For each of them $s(x)\ge\epsilon/4$.
\end{proof}

\begin{theorem}[A universal codebook]\label{fc:cover-new}
A single computable decoder recognizes every legal selector cut on
$n$-bit strings with zero error using $n^2/4+O(n)$ ordinary advice bits
and at most $n$ nonadaptive Boolean comparisons. More generally, suppose
that a cover of a specified $K$-vertex subtournament suffices to recognize
the cut. With $v=\log K$ and $u=\min\{v,n/2\}$, the decoder uses
$u(n-u)+O(n)$ advice bits and at most $\lceil v\rceil+1$ comparisons.
\end{theorem}
\begin{proof}
Write $S$ for the side to be covered, reversing the orientation if
necessary. Assume $K\ge2$. Apply Lemma~\ref{fc:fixedpoint} with
$\epsilon=1/(32n)$ and $\zeta=2/3$, and put $r=\lceil J\rceil$.
Since $(v+1)\log(1+1/(32n))<\log(4/3)$ for $v\le n$, the lemma gives
$J<v+1$ and $r\le\lceil v\rceil+1$. Also $J\epsilon=O(1)$, so its
power-moment bound gives $-J\log A\ge J-O(1)$.

Let $\pi=\nu^{\otimes r}$ and let $\mathcal C$ be its covering tuples.
A union bound gives $\pi(\mathcal C)\ge1-\sum_xs(x)^r\ge1/3$.
H\"older's inequality applied to the restriction of $\pi$ to $\mathcal C$
yields
\[
 |\mathcal C|\ \ge\
 \frac{\pi(\mathcal C)^{J+1}}
      {(\sum_{t\in S^r}\pi(t)^{1+1/J})^J}
 \ \ge\ (1/3)^{J+1}A^{-rJ}
 \ \ge\ 2^{r^2-O(r+1)} .
\]
For a singleton take $r=1$. A uniform ambient tuple therefore covers
$S$ using anchors in $S$ with probability at least
$2^{-r(n-r)-C(r+1)}$ for one absolute integer $C$.

For each $r\le n+1$ sample a separate list of
$2^{2n+\max\{0,\,r(n-r)+C(r+1)\}}$ independent ambient tuples.
For each marked tournament take a qualifying $r$. Its failure
probability is at most $\exp(-2^{2n})$. The number of tournaments
with a marked vertex subset is at most
$2^{\binom{2^n}{2}+2^n}$, so one choice of lists works for every
marked tournament at once. The lexicographically first such family
is computable by finite enumeration. For every nonempty marked subset verify
that some list with $r\le\lceil\log K\rceil+1$ covers it using anchors in it.
The decoder regenerates the indexed tuple and compares the input
with these anchors in the indicated orientation. An advice flag
handles an empty side. Apart from $O(n)$ bits the index costs
$r(n-r)$, bounded by $u(n-u)+O(n)$ for $r\le\lceil v\rceil+1$. For a nontrivial legal cut take the smaller side, for which $K\le 2^{n-1}$ and $r\le n$. The computation of the lists has unrestricted
running time.
\end{proof}

\section{Oblivious search for an equilibrium}\label{sec:games}
A symmetric constant-sum game is specified by an $N$-by-$N$ matrix
$A\in[0,1]^{N\times N}$ with $A_{ij}+A_{ji}=1$. On the action pair
$(i,j)$ the payoffs are $(A_{ij},A_{ji})$. In particular $A_{ii}=1/2$.
Subtracting $1/2$ from each payoff yields the equivalent zero-sum
normalization. We use the ordinary additive $\epsilon$-Nash condition: neither player
gains more than $\epsilon$ by unilateral deviation. An oblivious procedure
samples a pair of mixed strategies from a distribution that depends only
on $N$ and $\epsilon$. Let $P_\epsilon(N)$ be the supremum over such
distributions of the infimum, over games, of the probability of proposing an
$\epsilon$-Nash equilibrium. Let $L_\epsilon(N)$ be the minimum cardinality of
a game-independent list of profiles containing an $\epsilon$-Nash equilibrium
for every such game. All logarithms are to base two.

\begin{theorem}\label{g:main}
Fix $0<\epsilon<1/2$, put $n=\log N$, $\beta=1/2+\epsilon$, and set
\[
 c=1-H_2(\beta)
   =\beta\log(2\beta)+(1-\beta)\log(2(1-\beta)).
\]
Then
\[
 \frac{n^2}{4c}-O_\epsilon(n\log n)
 \ \le\ -\log P_\epsilon(N)
 \ \le\ \log L_\epsilon(N)
 \ \le\ \frac{n^2}{4c}+O_\epsilon(n).
\]
A proposal attaining the upper bound on $-\log P_\epsilon(N)$ chooses a
length uniformly from $1,\ldots,\lceil(n+1)/c\rceil$, draws that many
independent uniform actions, and uses their empirical distribution for
both players. The lower bound already holds when every off-diagonal
payoff belongs to $\{0,1\}$, and allows arbitrary pairs of mixed strategies.
\end{theorem}

For a mixed strategy $z$ define $v_A(z)=\max_i(Az)_i$.
The profile $(z,z)$ has expected payoff $1/2$ for each player and is
an $\epsilon$-Nash equilibrium exactly when $v_A(z)\le\beta$. The same
test controls asymmetric proposals. If $(x,y)$ is an $\epsilon$-Nash
equilibrium and $u=x^{\mathsf T}Ay$, its two deviation constraints imply
\[
 v_A(y)\le u+\epsilon,\qquad v_A(x)\le 1-u+\epsilon,\qquad
 v_A\big((x+y)/2\big)\le\beta .
\]
These implications use $A+A^{\mathsf T}=\mathbf1\mathbf1^{\mathsf T}$.

\begin{lemma}[Counting empirical strategies]\label{g:count}
For every such $A$ some integer $1\le r\le\lceil(n+1)/c\rceil$ has
at least $2^{c r^2-(c+2)r-1}$ ordered $r$-tuples of actions whose
empirical strategy $z$ satisfies $v_A(z)\le\beta$.
\end{lemma}
\begin{proof}
Let $\lambda=\log(\beta/(1-\beta))>0$. For a distribution $p$ on the
actions define
\[
 q_p(i)=2^{-\lambda\beta}\bigl(1+(2^\lambda-1)(Ap)_i\bigr).
\]
Since $p^{\mathsf T}Ap=1/2$, for every $p$ we have
\[
 \sum_i p_iq_p(i)=2^{-\lambda\beta}(1+2^\lambda)/2=2^{-c}.
\]
Lemma~\ref{fp:scores}, with $g_i=q_p(i)$ and $\zeta=1/2$, gives $p$
and $0<J\le(n+1)/c$ such that
\[
 \sum_i q_p(i)^J=1/2,\qquad p_i=2q_p(i)^J,\qquad
 \sum_i p_i^{1+1/J}=2^{1/J-c}.
\]
The final equality uses the exact mean $2^{-c}$ above. At this $p$ every
$q_p(i)$ is less than one, although the score of a profitable deviation
can exceed one at other distributions.

Set $r=\lceil J\rceil$. Sample $a_1,\ldots,a_r$ independently from $p$.
For any pure deviation $i$, convexity gives
$2^{\lambda u}\le1+(2^\lambda-1)u$ for $0\le u\le1$. Exponential
Markov and independence thus give
\[
 \Pr\left[\frac1r\sum_{j=1}^r A_{i a_j}>\beta\right]\le q_p(i)^r .
\]
Summing over $i$ bounds the failure probability by $Z=1/2$.
Let $\mathcal C$ be the good ordered tuples and $\mu=p^{\otimes r}$.
H\"older's inequality yields
\[
 |\mathcal C|\ge
 \frac{\mu(\mathcal C)^{J+1}}
 {(\sum_{\mathbf a}\mu(\mathbf a)^{1+1/J})^J}
 \ge 2^{crJ-r-J-1}\ge 2^{cr^2-(c+2)r-1}.
\]
\end{proof}

\begin{proof}[Upper bounds in Theorem~\ref{g:main}]
For the length in Lemma~\ref{g:count}, a uniform ambient tuple is good
with probability at least $2^{-r(n-cr)-O(r+1)}$. Since
$r(n-cr)\le n^2/(4c)$ and $r=O_\epsilon(n+1)$, sampling the length as
well gives success at least $2^{-n^2/(4c)-O_\epsilon(n)}$. An explicit
payoff matrix permits verification of a symmetric proposal in $O(Nr)$
arithmetic operations.

For a finite list that works for every real payoff matrix, take an even
integer $D=2\lceil n\rceil$ and use $\epsilon'=\epsilon-1/D$, positive
for all sufficiently large $N$. Round each game on unordered pairs to a
complementary pair of multiples of $1/D$, giving entrywise error at most
$1/D$ and preserving the diagonal. There are at most
$G=(D+1)^{N(N-1)/2}$ rounded games. Let $p_*$ be the guaranteed success
of the sampler above for $\epsilon'$. The constants in its estimate are
bounded uniformly for $\epsilon'\in[\epsilon/2,\epsilon]$ by the
displayed finite counting lemma. Smoothness of $1/c$ on this interval
gives $-\log p_*\le n^2/(4c)+O_\epsilon(n)$. A list of
$\lceil(\ln G+1)/p_*\rceil$ independent proposals works for every
rounded game with positive probability. Any successful symmetric proposal
has $v_A(z)\le 1/2+\epsilon$ in the original game. The covering list
adds $O(n)$ to the log size. Finally, uniform sampling from any
universal list gives $P_\epsilon(N)\ge1/L_\epsilon(N)$.
\end{proof}

The converse is proved in Theorem~\ref{g:probe}. The hard tournament has
one distinguished action per bucket. To be secure, a proposed strategy
must locate enough of these actions to prevent a fresh core row from
winning too much of its mass. A weighted binomial estimate handles
proposals with unequal weights.

\begin{corollary}[Correlated proposals]\label{g:cce}
The same bounds hold for obliviously proposing a coarse correlated
equilibrium of additive error $\epsilon$, and for a correlated equilibrium
with maximum expected gain at most $\epsilon$ under a swap of the
recommended action.
\end{corollary}
\begin{proof}
Let an arbitrary joint proposal have marginals $x,y$, and let the row
player's expected payoff be $u$. The two coarse deviation inequalities
again give $v_A(y)\le u+\epsilon$ and $v_A(x)\le1-u+\epsilon$.
Thus such a proposal yields an $\epsilon$-secure $(x+y)/2$ by a
game-independent map. For the converse, take the product $z\otimes z$
of an $\epsilon$-secure strategy with itself. The opponent's action
is independent of one's recommendation, so the best swap rule earns
at most $v_A(z)$. Following the recommendation earns $1/2$.
\end{proof}

In a tournament game the entries of $z$ can be interpreted as a lottery
over alternatives in pairwise comparisons. Security means that no single
alternative defeats this lottery with probability greater than
$1/2+\epsilon$. This is the approximate von Neumann winner criterion used
in preference learning~\cite{DHKSSZ15}.

\section{Consequences for tournaments}\label{sec:tournaments}
The quarter in Theorem~\ref{fc:cover-new} is not specific to selectors.
This section records three consequences of Lemma~\ref{fc:fixedpoint} for
tournaments alone. A set $D$ of vertices \emph{dominates} a tournament if
every vertex outside $D$ is beaten by a member of $D$, and a tuple
dominates if the set of its entries does. Reversing all edges turns
dominating sets into the covers of Section~\ref{fc:covers}, so
Lemma~\ref{fc:fixedpoint} and Theorem~\ref{fc:cover-new} apply to the
reversed tournament. Write $\gamma(T)$ for the domination number of $T$.

\subsection{Many dominating tuples}
\begin{corollary}[Many dominating tuples]\label{cor:many-absorbing}
Every tournament $T$ on $K\ge2$ vertices has a distribution $\nu$ on its
vertices, a real $J>0$ and an integer $r=\lceil J\rceil\le\lceil\log K\rceil+1$
with the following properties: $r$ independent samples from $\nu$
dominate $T$ with probability at least $1/3$, and the R\'enyi entropy of
order $1+1/J$ of $\nu$ is at least $J-O(1)$. In particular $T$ has at
least $2^{r^2-O(r)}$ dominating $r$-tuples, and $r\ge\gamma(T)$.
\end{corollary}
\begin{proof}
Apply Lemma~\ref{fc:fixedpoint} to the reversed tournament with
$\zeta=2/3$ and $\epsilon=1/(32(\lceil\log K\rceil+1))$, and argue as in
the proof of Theorem~\ref{fc:cover-new} with $\lceil\log K\rceil$ in place
of $n$. The R\'enyi entropy of order $1+1/J$ is $-J\log A$. A dominating
$r$-tuple exists, so $r\ge\gamma(T)$.
\end{proof}

The count is not a consequence of the existence of one small dominating
set. A tournament can have a unique minimum dominating set, and padding one
dominating set of size $\gamma$ with arbitrary vertices gives only about
$K^{r-\gamma}$ dominating $r$-tuples. For tournaments with
$\gamma(T)\ge\log K-2\log\log K-O(1)$, which exist by
Erd\H{o}s's argument~\cite{Erd63}, padding at a scale
$r\le\lceil\log K\rceil+1$ gives at most $2^{O(\log K\log\log K)}$ tuples,
while Corollary~\ref{cor:many-absorbing} gives
$2^{r^2-O(r)}\ge2^{\log^2K-O(\log K\log\log K)}$.

\begin{remark}[Why not a minimax strategy]\label{rem:minimax}
An optimal strategy $p^*$ of the symmetric zero-sum game on $T$ is beaten
by no vertex on average, so each vertex escapes a $p^*$-random dominator
with probability at most $1/2$, and $\lceil\log K\rceil+2$ samples from
$p^*$ dominate with probability at least $3/4$. Such a strategy can have
very small support. Let $A$ be a regular tournament on $m=2\lceil\log
K\rceil+1$ vertices, and let $B$ carry a uniformly random tournament on
$K-m$ further vertices, where each $x\in B$ is beaten by an independent
uniformly random set of $(m+1)/2$ vertices of $A$ and beats the rest. The
uniform distribution on $A$ is optimal, and its entropy is
$\log\log K+O(1)$. A union bound over sets of size
$d=\lfloor\tfrac14\log_3K\rfloor$ shows that with high probability
$\gamma(T)>d$: for each such set $D$, each $x\in B\setminus D$ escapes $D$
independently with probability at least $3^{-d}$. So $r$ samples from
$p^*$ produce at most $m^r=2^{O(\log K\log\log K)}$ different tuples at a
scale where Corollary~\ref{cor:many-absorbing} gives $2^{\Omega(\log^2K)}$.
The finite exponent $J$ in Lemma~\ref{fc:fixedpoint} is what forces the
spread.
\end{remark}

\subsection{Random domination}
How likely is a random set of $t$ vertices to dominate an $N$-vertex
tournament? A random tournament is dominated by a random set of size
$\log N+O(\log\log N)$ with high probability. The worst case is much
smaller, and its exponent has the same form $r(L-r)$ as the advice cost,
with $L=\log(N/t)$ in place of $n$.

\begin{theorem}[Random domination]\label{thm:random-dom}
There is an absolute constant $C$ with the following properties. Let
$N\ge2$ and $\lceil\log N\rceil+1\le t\le N$, and put $L=\log(N/t)$.
\begin{enumerate}
\item In every tournament on $N$ vertices, a uniformly random $t$-subset
dominates with probability at least $2^{-L^2/4-C(L+1)\log(L+2)}$.
\item If $L\ge C$, some tournament on $N$ vertices is dominated by a
uniformly random $t$-subset with probability at most
$2^{-L^2/4+C(L+1)\log(L+2)}$.
\end{enumerate}
\end{theorem}
\begin{proof}
\emph{Sampling with replacement suffices.} Draw $t$ independent uniform
vertices and complete the set of distinct draws by uniformly random new
vertices to a $t$-set. The resulting distribution on $t$-sets is invariant
under permutations of the vertices, hence uniform, and it contains the
drawn set. Domination is monotone, so it suffices to bound the probability
that the drawn set dominates. Let $C_0$ be the constant in the count
$2^{r^2-C_0r}$ of Corollary~\ref{cor:many-absorbing}.

\emph{Few samples.} Suppose $t<4(\lceil\log N\rceil+2)$. The first $r$ draws
form a uniform $r$-tuple with $r\le\lceil\log N\rceil+1\le t$, so they
dominate with probability at least
$2^{r^2-C_0r}/N^r=2^{-r(\log N-r)-C_0r}\ge2^{-\log^2N/4-C_0r}$. Here
$\log N=L+\log t$ and $\log t=O(\log(L+2))$: either $\log N<64$ and all
quantities are bounded, or $\log(4\log N+12)\le\tfrac12\log N$, whence
$\log N\le2L$. So $\log^2N/4\le L^2/4+O((L+1)\log(L+2))$.

\emph{Many samples.} Suppose $t\ge4(\lceil\log N\rceil+2)$. Split the draws
into a first half of $t_1=\lfloor t/2\rfloor\ge2t/5$ and a second half of
$t_2=\lceil t/2\rceil$. Let $S'$ be the set of vertices that are neither
drawn in the first half nor beaten by a first-half draw. A vertex $x$ with
$d^-(x)$ in-neighbours lies in $S'$ with probability at most
$\exp(-t_1(d^-(x)+1)/N)$. The $k$ vertices of smallest in-degree span a
tournament with average in-degree $(k-1)/2$, so the $k$-th smallest
in-degree is at least $(k-1)/2$, and
\[
 \mathbb E|S'|\le\sum_{k\ge1}e^{-t_1(k+1)/(2N)}\le\frac{2N}{t_1}\le\frac{5N}{t}.
\]
By Markov's inequality $K'=|S'|\le10N/t$ with probability at least $1/2$.
Fix such a first half. If $K'=0$ the first half dominates. Otherwise
Corollary~\ref{cor:many-absorbing} applied to the subtournament on $S'$
(trivially when $K'=1$) gives $r'\le\log K'+2\le L+6$ and at least
$2^{r'^2-C_0r'}$ dominating $r'$-tuples of $S'$. Every vertex outside $S'$
is drawn or beaten in the first half, so the draws dominate $T$ whenever
the second half contains a dominating tuple of the subtournament on $S'$.
Let $X$ be the number of second-half draws in $S'$; it is binomial with
parameters $t_2$ and $q=K'/N$. Given $X\ge r'$, the first $r'$ of them are
independent and uniform on $S'$, so they dominate the subtournament with
probability at least $2^{-r'(\log K'-r')-C_0r'}$.

If $t_2q\ge r'$, the median of $X$ is at least $r'$, and the overall
probability is at least
\[
 \tfrac14\,2^{-r'(\log K'-r')-C_0r'}\ge2^{-(L+4)^2/4-O(L+1)}.
\]
Otherwise $q<r'/t_2\le1/2$, because $r'\le\log N+2$ and
$t_2\ge2(\log N+2)$. Then, using $(1-q)^{t_2}\ge e^{-2qt_2}$,
\[
 \Pr[X\ge r']\ge\binom{t_2}{r'}q^{r'}(1-q)^{t_2}
 \ge\Bigl(\frac{t_2q}{r'}\Bigr)^{r'}e^{-2r'},
\]
and the product with the previous bound is at least
\[
 \Bigl(\frac{t_2K'}{Nr'}\Bigr)^{r'}
 2^{-r'\log K'+r'^2-O(r')}
 =2^{-r'(\log(N/t_2)-r')-r'\log r'-O(r')}
 \ge2^{-(L+1)^2/4-O((L+1)\log(L+2))},
\]
since $\log(N/t_2)\le L+1$ and $r'\le L+6$. The factor $K'$ cancels.
Multiplying by $1/2$ for the first half proves the first assertion.

\emph{The construction.} Let $m=\lfloor L/2\rfloor$, $K=2^m\le N$ and
$d=\lfloor m-2\log m\rfloor$. A uniformly random tournament $T_0$ on $K$
vertices has $\gamma(T_0)\ge d$ with positive probability~\cite{Erd63}:
for $s=d-1$ the expected number of dominating $s$-sets is at most
$K^s\exp(-(K-s)2^{-s})\le2^{m^2}e^{-m^2}<1$, since
$(K-s)2^{-s}\ge(K/2)\cdot2m^2/K$. Place such a $T_0$ on a set $C$ of $K$
vertices, let every vertex of $C$ beat every vertex outside $C$, and
orient the remaining edges arbitrarily. Only vertices of $C$ beat vertices
of $C$, so every dominating set $D$ meets $C$ in a dominating set of
$T_0$, and $|D\cap C|\ge d$. A fixed $d$-set lies in a uniform $t$-subset
with probability at most $(t/N)^d$, so
\[
 \Pr[\,|D\cap C|\ge d\,]\le\binom Kd\Bigl(\frac tN\Bigr)^d
 \le\Bigl(\frac{Kt}N\Bigr)^d=2^{-d(L-m)}.
\]
Finally
$d(L-m)\ge m(L-m)-(2\log m+1)L\ge L^2/4-1/4-(2\log L+1)L$.
\end{proof}

The same two estimates determine the size of universal families. Let
$U(N,t)$ be the least number of $t$-subsets of an $N$-set such that every
tournament on it is dominated by one of them.

\begin{corollary}[Universal dominating families]\label{cor:universal-dom}
Under the assumptions of Theorem~\ref{thm:random-dom}, with $L\ge C$,
\[
 \frac{L^2}4-C(L+1)\log(L+2)\ \le\ \log U(N,t)\ \le\
 \frac{L^2}4+C(L+1)\log(L+2)+2\log N.
\]
\end{corollary}
\begin{proof}
For the upper bound take $N^2\,2^{L^2/4+C(L+1)\log(L+2)}$ independent
uniform $t$-subsets. A fixed tournament is missed by all of them with
probability at most $e^{-N^2}$, and there are fewer than $e^{N^2}$
tournaments. For the lower bound place the construction of
Theorem~\ref{thm:random-dom} on a uniformly random $K$-set $C$. A fixed
$t$-set $F$ meets $C$ in at least $d$ vertices with probability at most
$\binom td(K/N)^d\le2^{-d(L-m)}$, so fewer than $2^{d(L-m)}$ sets miss
some placement.
\end{proof}

The upper bound is dominated by the term $L^2/4$ once
$L\gg\sqrt{\log N}$, for instance for every $t\le N^{1-\delta}$ with
$\delta>0$ fixed.

\subsection{Description length of dominating sets}
Every tournament on $\bits^n$ has a dominating set of at most $n+1$
vertices, which takes about $n^2$ bits to write down. Write $C(D\mid n)$
for the Kolmogorov complexity of a set $D$, given $n$ and encoded as a
sorted list.

\begin{theorem}[Describing dominating sets]\label{thm:kolmogorov-dom}
\leavevmode
\begin{enumerate}
\item Every tournament on $\bits^n$ has a dominating set $D$ of at most
$n+1$ vertices with $C(D\mid n)\le n^2/4+O(n)$.
\item For all large $n$ and every $s\ge1$ there is a tournament on
$\bits^n$ in which every dominating set $D$ with $|D|\le s$ satisfies
$C(D\mid n)\ge n^2/4-\tfrac n2\log s-O(n\log n)$. For $|D|$ bounded by a
polynomial in $n$, this is $n^2/4-O(n\log n)$.
\end{enumerate}
\end{theorem}
\begin{proof}
For the first part, the lists in the proof of Theorem~\ref{fc:cover-new}
are computable from $n$ and contain, for every tournament and every
nonempty vertex set, a covering tuple of length $r\le n+1$ inside that
set. Apply this to the reversed tournament and the whole vertex set. The
index of the tuple, together with $r$, costs $r(n-r)+O(n)\le n^2/4+O(n)$
bits, and a fixed program prints the tuple from it.

For the second part let $m=\lfloor n/2\rfloor$, $K=2^m$ and
$d=\lfloor m-2\log m\rfloor$, and place the construction of
Theorem~\ref{thm:random-dom} on a uniformly random $K$-set $C$. A fixed
$D$ with $|D|\le s$ meets $C$ in at least $d$ vertices with probability at
most $(sK/2^n)^d=2^{-d(n-m-\log s)}$. Fewer than $2^b$ sets have
$C(D\mid n)<b$. For $b=d(n-m-\log s)$ some placement of $C$ therefore
meets none of them in $d$ vertices, and then none of them dominates.
Finally $d(n-m-\log s)\ge n^2/4-\tfrac n2\log s-(2\log n+1)n-1$.
\end{proof}

The second part needs some bound on $|D|$: the whole vertex set dominates
and has complexity $O(1)$ given $n$.

\begin{remark}[Time-bounded descriptions]\label{rem:kpoly}
For tournaments computed by circuits of polynomial size, the method of
Section~\ref{fc:efficient} should give dominating sets whose
polynomial-time-bounded complexity is $n^2/4+O(n)$ under $\mathrm H$, with
the constant depending on the circuit size. In
this setting membership in the covered set is trivial. We have not
written out the details. This differs from language
compression~\cite{BLvM05}, which describes every element of a set $A$ in
about $\log|A|$ bits: here one element of a sparse set of tuples is
described in about $\log(\text{universe}/|A|)$ bits. A direct
hashing-plus-generator argument for such a set, which is defined by a
universal condition, would need hardness against circuits with
$\Sigma_2^{\mathrm P}$ gates.
\end{remark}

\subsection{Other covering relations}
\begin{remark}[Other covering relations]\label{rem:covering-games}
The tournament enters the first two parts of Lemma~\ref{fc:fixedpoint}
only through the identity
$\Pr_{x,a\sim p}[a\succ x]=(1-\sum_xp(x)^2)/2$. Suppose instead that a
finite set $S$ carries a relation in which an anchor $a$ may \emph{miss} a
vertex $x$, no anchor misses itself, and
$\Pr_{x,a\sim p}[a\text{ misses }x]\le\theta$ for every distribution $p$ on
$S$, where $1/2\le\theta<1$. The same proof gives
$J\le\log(K/\zeta)/\log(1/(\theta+\epsilon))$ and
$A\le\zeta^{-1/J}(\theta+2\epsilon)$. With $\epsilon$ a small multiple of
$\beta/\log K$, where $\beta=\log(1/\theta)$, the count in
Theorem~\ref{fc:cover-new} gives at least $2^{\beta r^2-O(r)}$ covering
$r$-tuples for some $r=O(\log K/\beta)$.

For example, let $G$ be an oriented graph on $S$ with independence number
$\alpha$, and say that $a$ covers $x$ when $x=a$ or $x\to a$. A $p$-random
anchor misses a $p$-random vertex with probability
$(1-\sum_xp(x)^2)/2$ plus the $p\otimes p$ mass of the unordered
nonadjacent pairs. The Motzkin--Straus theorem~\cite{MS65} bounds that mass
by $(1-1/\alpha)/2$, so $\theta=1-1/(2\alpha)$ works. Tournaments are the
case $\alpha=1$.
\end{remark}

\section{Polynomial-time decoding under circuit hardness}\label{fc:efficient}
Let H assert that some function in $\mathrm E$ requires circuits of size
$2^{c\ell}$ with SAT gates on all sufficiently large input lengths
$\ell$, for a constant $c>0$. Circuit size charges the query wires.
All logarithms below are binary.

\begin{theorem}\label{fc:efficient-quarter}
Under H, every $\mathrm P$-selective language has a deterministic
polynomial-time interpreter with $n^2/4+O_L(n)$ ordinary advice
bits. If its smaller nonempty side has size $K\le2^m$ for an integer $0\le m\le n-1$,
the interpreter uses at most $m+1$ nonadaptive selector
comparisons. Its advice length is bounded by
\[
 \max_{1\le r\le m+1}r(n-r)+O_L(n).
\]
\end{theorem}

\paragraph{Two facts about short descriptions.}
Orient the smaller side $S$ so that it beats its complement. Membership
in $S$ has polynomial-size nondeterministic circuits. Take a vertex
$w\in S$ of maximum indegree within $S$. Each $x\in S$ reaches $w$
in at most two steps, since otherwise all inneighbors of $w$, together
with $w$, beat $x$. No vertex outside $S$ reaches $w$.

We use two facts about short descriptions of strings. In each statement
the string to describe has length $N$, polynomially bounded in $n$, with
$N\ge n$. Only the generator and affine map in the second tool are used
by the reconstruction algorithm.
\begin{enumerate}
\item If $Y_a\subseteq\bits^d$ is recognized nondeterministically on
$(a,y)$ by a polynomial-size circuit, $d=\mathrm{poly}(n)$, and
$1/\eta=\mathrm{poly}(n)$, a polynomial-size nondeterministic circuit
accepts every $a$ with $|Y_a|\ge T$ and only $a$ with
$|Y_a|\ge(1-\eta)T$. For completeness, take $j=\lceil3/\eta\rceil$
and $b=\lceil\log(T^j)\rceil+1$ for $1\le T\le2^d$. A random
pairwise independent hash of $j$-tuples into $\bits^b$ hits a specified
target on $Y_a^j$ with probability at least $7/32$ in the large case
and at most $1/16$ if $|Y_a|<(1-\eta)T$. For the lower bound, apply
two-term inclusion--exclusion to $T^j$ elements, whose expected number
of hits is in $(1/4,1/2]$. For the upper bound use
$(1-\eta)^j<1/8$. Fix $O(N)$ independent hashes so the fraction of
hits, thresholded at $9/64$, separates these cases simultaneously
for every $N$-bit $a$. Chernoff's bound and a union bound ensure
existence. A witness lists successful hashes, preimages and their
membership witnesses. For $T>2^d$ use a rejecting circuit. This is
the nonuniform set lower-bound test of Goldwasser and Sipser~\cite{derand:GS}.
\item Under H, suppose $W\ne\varnothing$ is contained in the
acceptance set $W'\subseteq\bits^N$ of a nondeterministic circuit of size $n^{O(1)}$.
Some $a\in W'$ has a polynomial-time description of length
$N-\log|W|+O(\log n)$. The reconstruction algorithm depends only on a polynomial bound on the circuit size. To see this, fix
$b=\lceil N-\log|W|\rceil$. A uniform affine map
$F:\bits^N\to\bits^N$ takes the $2^b$ zero-padded $b$-bit strings
pairwise independently to uniform outputs. The hit count on $W$ has
mean at least one and variance at most its mean, so the probability
of a hit is at least $1/2$. Whether some index is mapped into $W'$
is an NP test of the map, implementable by a polynomial-size circuit
with one SAT gate. The Klivans--van Melkebeek generator under H
with logarithmic seed fools this test to error $1/4$
\cite{derand:KvM,derand:Zim}. Encode the seed, index length, and hit
index. Decoding evaluates the generator and affine map in polynomial
time.
\end{enumerate}

\begin{proof}[Proof of Theorem~\ref{fc:efficient-quarter}]
An empty side needs a flag. For $K<n$, greedy halving gives a cover
with $r\le\lfloor\log K\rfloor+1$ anchors. Listing it costs
$rn=r(n-r)+O(n)$ since $r^2=O(\log^2 n)=O(n)$. For bounded $n$ we may likewise use a greedy cover by adjusting the
constant in the advice bound. Assume $K\ge n$ and $n\ge2048$. Apply
Lemma~\ref{fc:fixedpoint} with $\epsilon=1/(32n)$ and $\zeta=2/3$.
As in Theorem~\ref{fc:cover-new}, $r=\lceil J\rceil\le m+1$, and
$A=\sum_x\nu(x)^{1+1/J}$ satisfies $-J\log A\ge J-O(1)$.

\emph{Finite weights.}
Let $\sigma=(\epsilon/4)^3/2$, $\tau=\sigma/(64r)$ and $M=n/\tau^2$.
These choices make $M$, $M\tau$ and $M\sigma$ integers.
Hoeffding's bound and a union bound give $M$ anchors sampled from
$\nu$ whose empirical miss frequencies $\widehat s(x)$ approximate
$s(x)$ within $\tau$ simultaneously for $x\in S$. Set
\[
 t(x)=\max\{\sigma,\widehat s(x)+\tau\},\quad
 \widehat p(x)=(1-\epsilon)\widehat s(x)^J/\zeta+\epsilon/K,\quad
 h(x)=\max\{0,\lfloor\log(1/\widehat p(x))\rfloor\}.
\]
Write $R=\lfloor\log(K/\epsilon)\rfloor$, so $0\le h\le R$. The sampled names
and the tables for the integer values $Mt(x)$ and $h(x)$ can be
hardwired into polynomial-size circuits. There are only $M+1$ possible
frequencies. These tables specify integer outputs even when $J$ is
irrational. The following estimates hold:
\[
 s(x)\le t(x),\qquad \sum_{x\in S}t(x)^r\le0.70,\qquad
 \nu(x)\le2^{1-h(x)},\qquad h(x)\ge\log(1/\nu(x))-2.
\]
Indeed, where $s(x)\ge\sigma$ the relative error in $\widehat s(x)$
is at most $1/(64r)$ and $t(x)/s(x)\le1+1/(32r)$. The corresponding
terms in the power sum are at most $e^{1/32}s(x)^J$, and the
estimated and true values of $\nu(x)$ differ by at most a factor two.
For $s(x)<\sigma$ use $t(x)\le2\sigma$ and
$(2\sigma)^J=(\epsilon/4)^{3J}\le(4\zeta/K)^3$.
The resulting full sum is at most $e^{1/32}\zeta+K(4\zeta/K)^3<0.70$
for large $n$. The small terms' individual
nonuniform mass before smoothing is at most $\epsilon/K$, both in
the empirical and the true law, since $(4\zeta/K)^3/\zeta\le\epsilon/K$ for $K\ge n\ge2048$. The factor-two comparison then gives
$h(x)\ge\log(1/\nu(x))-2$ and $\nu(x)\le2^{1-h(x)}$.

\emph{Paying for the anchor descriptions.}
For $\rho=\mathbb E_\nu 2^{-h(a)/J}$, rounding the weights costs at most
two bits per anchor in the power-moment estimate.
\[
 \rho\le 2^{2/J} A,\qquad -J\log\rho\ge J-O(1).
\]

\emph{Anchor blocks.}
Describing the anchors separately would spend $O(\log n)$ bits per
anchor on seeds and length fields. We describe whole blocks. A block of
$k$ anchors can cost a factor $(\min_x t(x))^{-k}$ in the accuracy of
the counting test. The following choice keeps that factor polynomial
and the total description overhead linear. Set
\[
 j=\min\!\left\{\left\lceil\frac{n+6}{r}\right\rceil,
                 \left\lceil\log(1/\sigma)\right\rceil\right\},
 \qquad w(x)=\max\{t(x),2^{-j}\},\qquad
 b=\max\!\left\{1,\left\lfloor\frac{\log n}{j}\right\rfloor\right\}.
\]
If the second term attains the minimum defining $j$, then $w=t$.
Otherwise $K2^{-jr}\le 1/64$. Hence $\sum_xw(x)^r\le 0.70+1/64$.
Partition the $r$ positions into $B=\lceil r/b\rceil$ consecutive
blocks, the last possibly shorter. For every block length $k$,
\[
 2^{jk}=n^{O(1)},\qquad B\log n=O(n).
\]
Indeed, $jk\le\max\{j,\log n\}=O(\log n)$, and
$1/b\le 2j/\log n$ while $rj\le n+6+r$. Put $D=M2^j$.
All $D w(x)$ are integers and $D=n^{O(1)}$.

Let $S_i$ consist of the vertices of $S$ still uncovered after
$i$ chosen anchors. At the block boundaries use the potential
\[
 \Phi_i=\sum_{x\in S_i}w(x)^{r-i},\qquad
 \Xi_i=\tfrac1{16}\,2^{-\sum_{u\le i}h(a_u)/J}\rho^{-i},
 \qquad P_i=\Phi_i+\Xi_i.
\]
Initially $P_0\le 0.70+1/64+1/16$. Choose a provisional block
$\mathbf a=(a_{i+1},\ldots,a_{i+k})$ from $\nu^{\otimes k}$.
A surviving vertex $x$ remains uncovered with probability $s(x)^k$,
so $s(x)\le w(x)$ gives $\mathbb E\Phi_{i+k}\le\Phi_i$. Independence
gives $\mathbb E\Xi_{i+k}=\Xi_i$. With $\delta=1/(16r)$, Markov's
inequality therefore assigns mass at least $\delta/(1+\delta)$ to
blocks with $P_{i+k}\le(1+\delta)P_i$. Bin these blocks by
$H=\sum_{u=i+1}^{i+k}h(a_u)$. There are at most $kR+1$ bins, and each
tuple in bin $H$ has mass at most $2^{k-H}$. Some bin thus contains
a set $V$ of such blocks of cardinality
\[
 |V|\ge 2^{H-k}\frac{\delta}{(1+\delta)(kR+1)} .
\]

\emph{Certifying a block.}
Fix this bin, put $e=r-i-k$, $\Phi'_i=\sum_{x\in S_i}w(x)^e$, and
write $\xi$ for the common value of $\Xi_{i+k}$ in the bin. If
$W(\mathbf a)$ is the sum of $w(x)^e$ over the vertices of $S_i$
certified by at least one entry of $\mathbf a$, then
\[
 P_{i+k}=\Phi'_i+\xi-W(\mathbf a),\qquad
 T_0=\Phi'_i+\xi-(1+\delta)P_i .
\]
Every block in $V$ has $W(\mathbf a)\ge T_0$. In particular
$\xi\le(1+\delta)P_i$, so $T_0\le\Phi'_i\le2^{jk}P_i$.

The integer $D^e W(\mathbf a)$ counts pairs $(x,z)$ for which $x\in S_i$,
some block entry certifies $x$, and $1\le z\le(Dw(x))^e$.
This is a nondeterministically recognizable set. Membership in $S$ uses
the path to the fixed king, and the earlier anchors determine $S_i$.
The binary representation of $z$ uses $O(n\log n)$ bits. Check that
each block entry is in $S$ and that the entries have total bin value $H$.
When $T_0>0$, apply the set lower-bound test at integer threshold
$\lceil D^eT_0\rceil$ with relative slack $\eta=\delta\,2^{-jk}$.
Its inverse is polynomial in $n$. All of $V$ pass. Any passing block has
\[
 W(\mathbf a)\ge(1-\eta)T_0\ge T_0-\delta P_i,
 \qquad P_{i+k}\le(1+2\delta)P_i .
\]
When $T_0\le0$, the membership and bin tests alone suffice.
The earlier anchors, empirical tables, king and integer threshold occur
only in this nonuniform test. Inductively $P_i<1$, so the bound on
$T_0$ gives a threshold of $O(n\log n)$ bits.

Apply the second short-description tool with $N=kn$. Some passing
block has description length
$kn-\log|V|+O_L(\log n)\le kn-H+k+O_L(\log n)$.
Choosing such a block at every step keeps
$P_i\le (1+2\delta)^B P_0\le e^{1/8}(0.70+1/64+1/16)<1$ at each
boundary. At the last boundary $\Phi_r=|S_r|$ is an integer and hence
zero. Also $\Xi_r<1$ gives
\[
 \sum_{u=1}^r h(a_u)>-Jr\log\rho-4J\ge r^2-O(r).
\]
The decoder reconstructs all $r$ names from the block descriptions,
compares the input with them in the indicated orientation, and complements
the outcome for a negative side. Framing the descriptions takes at most
$O(\log n)$ bits per block. Summing over the blocks gives
\[
 rn-\sum_{u=1}^r h(a_u)+O_L(r+B\log n)
       \le r(n-r)+O_L(n).
\]
The reconstruction uses polynomial time and at most $r\le m+1$
nonadaptive selector comparisons.
\end{proof}

\section{The algebraic lower bound}\label{sec:alg-lower}
Write $K=2^m$, $a=n-m$, and $T=2^a$. Partition the $n$-bit vertices
into $K$ buckets of size $T$. In each bucket $\beta$ choose a uniform
core $t_\beta$ and a fair label $\lambda_\beta$, independently.
The members are exactly the cores of label one. Cores beat noncores,
positive cores beat negative cores, and equal-label cores use one
independent fair coin per unordered pair. Order noncores lexicographically
and set the diagonal to zero. This is the Boolean hidden-core distribution
of~\cite{BD}. Conditional on the table $E$, sample one uniform extension
from its polynomial fiber in each available finite field, independently
across fields. Degrees are bounded in each variable by a prescribed
integer at least two. All executions see the same sampled tuple $z$.
We first fix a deterministic interpreter $D$ making at most $q$ queries
on every history. Its description, tape when present, and advice word are
fixed independently of the draw. The next subsection defines visibility
and the addressed set $\Adr$ and proves their locality properties.

\subsection{Which entries a degree-two query can reveal}\label{sec:visibility}
Fix a table $E$ and, independently in each field, sample its extension uniformly
from the finite affine space of polynomials with individual degree at most $d$
agreeing with $E$ on the cube. Here $d\ge2$; it may depend on the field.
Fix the tape and advice of a classical interpreter for the following two lemmas.
A recorded transcript includes the queries and their answers.
For a set $S$ of queried points in one field, call a Boolean position
$(u,v)$ visible if no polynomial in the permitted degree class vanishes
on $S$ and restricts on the cube to the point indicator
$\delta_{(u,v)}$. Write $B(S)$ for these positions.
\begin{lemma}[Transcript privacy]\label{lem:visible}
The likelihood of a recorded transcript depends on the underlying table
only through the visible positions of its query set in each field.
\end{lemma}
\begin{proof}
For each invisible position choose a polynomial witnessing invisibility.
If two tables agree on $B(S)$, the sum of these polynomials weighted by
their entrywise differences translates one extension fiber onto the
other while preserving its evaluations on $S$. Uniform fiber measures
therefore give the same probability to every evaluation cylinder.
For a recorded adaptive transcript earlier answers determine later
queries. Its cylinder imposes just the recorded evaluation constraints,
which factor over fields conditional on the table.
\end{proof}

For points $P$ in one copy of field $n$-space, let $V(P)$ be the Boolean
vertices whose evaluation functionals on multilinear polynomials lie
in the span of the evaluations at $P$. Boolean evaluations form a dual
basis, so $|V(P)|\le |P|$. Use superscripts $(1),(2)$ for projections
onto the two copies of $n$-space.
\begin{lemma}[Product criterion]\label{lem:product}
For any $S=S_1\cup S_2$ and $(u,v)\in B(S)$, one has
$u\in V(S_1^{(1)})$ or $v\in V(S_2^{(2)})$.
\end{lemma}
\begin{proof}
If both alternatives fail, there are multilinear $g,g'$ vanishing on
$S_1^{(1)},S_2^{(2)}$, respectively, with $g(u)=g'(v)=1$.
The polynomial $\delta_u(Y)g(Y)\delta_v(Z)g'(Z)$ has individual
degree at most two, vanishes on $S$, and restricts on the cube to
$\delta_{(u,v)}$.
\end{proof}
For a family of query sets, let $\operatorname{Adr}$ be the union of
$V(S^{(1)})$ and $V(S^{(2)})$ over its fields. Both endpoints of a
visible position belong to $\operatorname{Adr}$, by the two covers
with an empty part. This monotone set has size at most twice the
number of queries.

\subsection{Discovery and errors}
\begin{lemma}[Discovery on one input]\label{lem:single}
Let $A_\beta$ count other cores in $\operatorname{Adr}$ of the
standalone run on $t_\beta$. If $T\ge4q$, then
$\Pr[A_\beta\ge h]\le(4eq/(hT))^h$ for every integer $h\ge1$.
\end{lemma}
\begin{proof}
Reveal the current core and label. Before answering each query, test
each newly addressed vertex in a fixed order, revealing its core
status and the labels and pair coins needed among revealed cores.
Lemma~\ref{lem:visible} permits sampling the next answer using only
the exposed table entries and fresh coins, by the ratio of successive
cylinder probabilities. Thus each undiscovered core remains uniform
among the untested tails of its bucket. Each of the at most $2q$ tests
discovers an additional core with conditional probability at most
$1/(T-2q)\le2/T$. Summing over the positions of $h$ successful tests
gives $\binom{2q}{h}(2/T)^h$.
\end{proof}

\paragraph{Many inputs to the same oracle.}
\begin{lemma}[Error count]\label{cor:saturation-good-rounds}
Let $Z$ count the core inputs on which $D$ errs. For $m\ge3$,
$T\ge32eq$, and $h\ge1$, put $p=2^{-h-1}$. Then
\[
 \Pr[Z\le pK/4]\le
 \left(\frac{32eq}{3T}\right)^{K/4}
 +8\left(\frac{8eq}{hT}\right)^h+\exp(-pK/16).
\]
\end{lemma}
\begin{proof}
Maintain the cumulative query sets and test every newly addressed
vertex. Choose the next input core from the least bucket whose core
location has not been exposed. Reveal that location, execute its
ordinary run, and defer its label and incident coins to the end of the
round. In each other unexposed bucket, either hitting its core or
accumulating $T/4$ distinct unsuccessful tests exposes its location,
label, and coins with already exposed cores except the current core.
Such a bucket receives no future round. Process tests before recording
the current answer, including when one query causes several tests.
Closing a round reveals its deferred variables for the exposed cores.

At the beginning of a round, its core $t_\beta$ was unaddressed in
the past. Its label and its coins are fresh. Indeed a fixed enriched history
specifies the source exposures, and its answer likelihood depends only
on the visible table values, all determined by those exposures.
Within a round the same assertion about other unexposed buckets
holds after conditioning on the current label and coins.
If a visible position now joins $t_\beta$ to $v$, apply
Lemma~\ref{lem:product} with the past and the current run as the two
parts (exchange them for the reversed position). Since the past does
not address $t_\beta$, the current run alone addresses $v$.

Fix a start-of-round history including $t_\beta$. Let $P_b$ be the
law of the ensuing history for current label $b$, excluding the
closing revelation. Under an auxiliary law $U$, the current core
loses to all positive cores and beats all negative cores. Keep all
other entries fixed and draw the extension fibers conditional on the
recorded past. The old cylinder has the same probability by privacy.
For a completed history $\tau$, let $d(\tau)$ count the other core
vertices joined \emph{to the current core} by a visible position in
the cumulative query sets. On fixing the outside variables, at most
$d(\tau)$ equal-label coins must take prescribed values for all visible
entries to agree with the auxiliary table. None was exposed during
the round. Transcript privacy and integration over the outside
variables therefore give $P_b(\tau)\ge2^{-d(\tau)}U(\tau)$.
Whichever output $D$ gives, its conditional error probability is at
least $2^{-h}U(d<h)$, including the prior $1/2$ on the opposite label.

Stop just after the tests of the first issued query for which $d\ge h$,
before its answer. Only the previous answered queries constrain coins
of the current core. Their incident count is at most $h-1$, even if
the last query increases $d$ by several units. The disjoint stopped
cylinders have $P_b$-probability at least $2^{-(h-1)}$ times their
$U$-probability. Current-run locality implies $A_\beta\ge h$ on each
such true history. Thus
\[
 U(d\ge h)\le2^{h-1}\Pr[A_\beta\ge h\mid\text{past}].
\]
Call a round good when the conditional probability on the right
before multiplication is at most $2^{-h}$. It errs with conditional
probability at least $p$. Give each bucket its genuine starting
history if selected and a cemetery symbol otherwise. By the tower
property, the probability that bucket $\beta$ has a bad round is
at most $2^h\Pr[A_\beta\ge h]$, where $A_\beta$ is defined by its
standalone run even if unselected. Markov's inequality and
Lemma~\ref{lem:single} bound the probability of at least $K/8$ bad
rounds by $8(8eq/(hT))^h$.

There are at most $2Kq$ tests, so at most $8Kq/T<K/8$ buckets
reach the unsuccessful-test threshold. Before then, a test naturally
discovers a fresh core with conditional probability at most $4/(3T)$.
The probability of at least $K/4$ natural discoveries is at most
$\binom{2Kq}{K/4}(4/(3T))^{K/4}\le(32eq/(3T))^{K/4}$.
Outside these two exceptional events there are at least $K/2$ good
rounds. Select them in chronological order, padding any missing
slots by independent Bernoulli-$p$ variables. Goodness is known
before the output. Each selected or padded error indicator $X$ has
conditional expectation at least $p$ at selection, so its conditional
exponential moment at $-\ln 2$ is at most $1-p/2$. Iterating over
$K/2$ slots bounds the probability of at most $pK/4$ selected errors
by $\exp(-pK/16)$. On the nonexceptional event these are actual errors.
\end{proof}

\subsection{Counting advice words}
\begin{theorem}[All-core success]\label{thm:allcore}
Let $D$ be a deterministic interpreter making at most $Q$ queries on every
history, fixed independently of the draw. Suppose $m\ge3$, $T=2^{n-m}\ge32eQ$
and $j=m-\lceil4\log(n+2)\rceil\ge1$. Then
\[
 \Pr[D\text{ is correct on every core input}]
 \le\left(\frac{32eQ}{3T}\right)^{K/4}
 +8\left(\frac{8eQ}{jT}\right)^j
 +\exp\!\left(-\frac{(n+2)^4}{32}\right).
\]
If $Q\le n^d$, the right side is at most $2^{-m(n-m)+O_d(n\log n)}$ for
large $n$.
\end{theorem}
\begin{proof}
Correctness on every core input means $Z=0\le pK/4$. Apply
Lemma~\ref{cor:saturation-good-rounds} with $h=j$, so $p=2^{-j-1}$.
Since $m-j\ge4\log(n+2)$, its last term is
$\exp(-2^{m-j-5})\le\exp(-(n+2)^4/32)$.
For the estimate, the middle term is
$2^{3-j(n-m)+j\log(8eQ/j)}$, and
$j(n-m)\ge m(n-m)-(4\log(n+2)+1)n$ while $j\log(8eQ)=O_d(n\log n)$.
The first term is at most $3^{-K/4}$ with $K\ge2^j(n+2)^4\ge n^4$.
Both the first and last terms are below $2^{-n^2}\le2^{-m(n-m)}$
for large $n$.
\end{proof}

\begin{corollary}[Ordinary advice]\label{cor:adaptive-advice}
Fix a randomized interpreter with a finite random tape independent of the
source, making at most $q=q(n)$ adaptive classical queries, $q$ polynomial.
Put $Q=(18(n+2)+3)q$. For large $n$ and every $m$ with
$\lceil4\log(n+2)\rceil+1\le m$ and $2^{n-m}\ge32eQ$, with probability at
least $1-2^{-n}$ over the hidden cores and the extension tuple, no ordinary
word of length at most $m(n-m)-C_qn\log n$ gives pointwise error $1/3$.
In particular, at $m=\lfloor n/2\rfloor$ ordinary advice at pointwise
error $1/3$ requires $n^2/4-O_q(n\log n)$ bits.
\end{corollary}
\begin{proof}
Repeat the machine independently with the same word an odd number of
times between $18(n+2)$ and $18(n+2)+2$ and take the majority. Pointwise
error $1/3$ drops below $2^{-n-2}$, so all $K\le2^n$ core inputs are
answered correctly on at least three quarters of the combined tapes.
For a fixed word $w$ and combined tape $R$, the amplified interpreter is a
deterministic $D_{w,R}$ with at most $Q$ queries. By Markov's inequality
and Theorem~\ref{thm:allcore}, the probability over the source that $w$
gives pointwise error $1/3$ is at most
$\tfrac43\,\mathbb E_R\Pr[D_{w,R}\text{ correct on every core}]
\le\tfrac43\,2^{-m(n-m)+O_q(n\log n)}$.
A union bound over the fewer than $2^{\ell+1}$ words of length at most
$\ell=m(n-m)-C_qn\log n$ finishes the proof for a suitable $C_q$. The words
may depend on the entire realized tuple of extensions.
\end{proof}

\begin{lemma}[Compressed exact cover]\label{lem:compressed-cover}
With probability $1-O(m^{-2})$ one deterministic polynomial-time
decoder is exact using $ma+O(n)$ ordinary bits and at most $m+2$
Boolean comparisons.
\end{lemma}
\begin{proof}
Take $\lceil\log m\rceil$ disjoint public groups of $4m$ buckets.
Each contains at least $m+2$ positives except with probability
$\exp(-\Omega(m))$. In each group choose its first $m+2$ positive
cores. Conditional on labels and locations, the probability that
all positives outside the groups are certified by a given candidate
is at least $1-2^m2^{-m-2}=3/4$, independently for the different
candidates. The probability all candidates fail externally is at
most $m^{-2}$. A union bound over candidates and positive vertices
inside the groups bounds the chance that any such vertex misses a
candidate by $4m\lceil\log m\rceil^2 2^{-m-2}$. Store a successful
group's index, its $4m$-bit selection mask, and the $m+2$ tails.
The decoder reconstructs its positive anchors in bucket order.
\end{proof}

\section{The multilinear threshold}\label{fc:degree}
A Boolean comparison reads one entry of a row of the table. In the
multilinear extension, evaluating at the midpoint of that row gives
its sum. The only advice needed to interpret this sum is the size of
the cut.

\begin{theorem}[The exact ordinary advice cost at degree one]\label{thm:mlcollapse}
Let $N=2^n$, $n\ge1$. Given the canonical multilinear extensions of a
selector table over the finite fields, a uniform deterministic
polynomial-time decoder recognizes every legal cut using $n+1$
fixed-width advice bits and $O(n/\log(n+1))$ nonadaptive queries over
prime fields of order at most $4n$. With a public prime $p>N$, one
query over $\F_p$ suffices, in time polynomial in $n+\log p$.

The fixed width $n+1$ is optimal even with the entire table available
and pointwise error strictly below $1/2$. If the length of a variable-length
word is supplied for free, the exact worst-case ordinary advice cost is $n$.
\end{theorem}
\begin{proof}
Write $k=|P|$. A member beats all $N-k$ nonmembers. A nonmember can
beat only the other nonmembers, of which there are at most $N-k-1$.
Consequently $x\in P$ exactly when $\deg^+(x)\ge N-k$.

For odd $p$, put $z_*=(2^{-1},\ldots,2^{-1})\in\F_p^n$. Every
Boolean Lagrange basis polynomial in the second argument has value
$2^{-n}$ at $z_*$. The canonical multilinear extension therefore obeys
\begin{equation}\label{eq:midpoint-degree}
 N E_{\rm ml}(x,z_*)=\deg^+(x)\pmod p .
\end{equation}
The residue determines the degree when $p>N$. For the uniform
small-field version, sieve all primes in $(2n,4n]$. Their product
exceeds $N$. Indeed, for $2n\ge21$, the interval contains more than
$3(2n)/(5\ln(2n))$ and fewer than $7(2n)/(5\ln(2n))$ primes
by Rosser and Schoenfeld~\cite[Corollary~3]{RS62}. Thus the natural
logarithm of their product exceeds $6n/5>n\ln2$. For $1\le n\le10$
the product inequality follows by direct multiplication. The Chinese
remainder theorem recovers the integer degree in $[0,N-1]$ from the
queried residues in polynomial time. A fixed-width word encodes
$k\in\{0,\ldots,N\}$ in $n+1$ bits. With variable lengths, encode
$k<N$ in $n$ bits and $k=N$ by the empty word.

For optimality fix one transitive tournament on $N$ ordered vertices.
Its $N+1$ initial segments are all legal cuts and have identical
oracle tables and identical canonical extensions. Two different cuts
disagree on some input. On that input a decoder using the same advice
has the same output distribution for both, so its error is at least
$1/2$ on one of them. Thus $N+1$ distinct words are necessary. There
are $2^\ell$ words of fixed width $\ell$ and $2^{\ell+1}-1$ of length
at most $\ell$, giving the two exact values.
\end{proof}

\section{The size of the components adjacent to a cut}
\label{focus:component}
The strong components of a tournament are linearly ordered, with all edges
pointing from earlier to later components. For a nontrivial legal cut
$(E,P)$, let $B^+$ be the last positive component and $B^-$ the first
negative component. Put $B=\min\{|B^+|,|B^-|\}$, $v=\log_2 B$, and
$u=\min\{v,n/2\}$.

\begin{theorem}[Advice at a prescribed boundary-component size]
\label{focus:component-profile}
There is a computable decoder using $u(n-u)+O(n)$ ordinary advice
bits and $\lceil v\rceil+1$ nonadaptive Boolean comparisons with zero error.
Conversely, fix any randomized decoder making polynomially many adaptive
classical algebraic queries with ordinary advice. For each sufficiently large
$n$ and each integer $3\le B\le 2^{n-1}$, there is a legal cut with
\emph{exactly} this boundary-component size and a fixed tuple of
individual-degree-two extensions requiring at least
$u(n-u)-O_M(n\log n)$ bits for pointwise error at most $1/3$.
The same additive estimate holds trivially for $B=1$; a tournament
component of size two does not occur. For a P-selective language a
deterministic polynomial-time decoder uses $n(\lfloor v\rfloor+1)+O(1)$
ordinary advice bits unconditionally. Under the SAT-gate hardness
hypothesis, the polynomial-time upper bound for a P-selective language is
$u(n-u)+O_L(n)$.
\end{theorem}

\begin{proof}
Orient the tournament so that the chosen boundary component $C$, of size
$B$, is the last component of its side and that side beats its complement.
A cover of $C$ by anchors in $C$ recognizes the entire side: all its
earlier components beat every anchor and all vertices on the other side
lose to every anchor. The ordinary cover-density bound on the tournament
induced by $C$ and the universal codebook construction give
$u(n-u)+O(n)$ bits by Theorem~\ref{fc:cover-new}. In detail, hardwire no instance data in the
codebook. Group tournaments with a marked subset by $\lceil\log_2 B\rceil$,
choose a codebook for each group by the finite probabilistic construction,
and let the advice specify the group, entry, prefix length and orientation.
There are at most $2^{O(2^{2n})}$ marked tournaments, which adds $O(n)$
index bits to minus the logarithm of the covering density.

For the unconditional polynomial-time bound run Ko's halving inside $C$.
In any nonempty $R\subseteq C$ some vertex is beaten by at least
$(|R|-1)/2$ others of $R$; as an anchor it certifies itself and them, so
fewer than $|R|/2$ vertices of $R$ remain. After $\lfloor v\rfloor+1$
anchors nothing remains. The advice lists these anchors in full together
with the orientation and side bits, and the decoder compares its input with
each anchor using the selector.

For the conditional upper bound one needs a small nondeterministic
membership test for $C$ in the choice proof. In the tournament induced
by $C$, take a vertex $a$ of maximum outdegree and a vertex $b$ of
maximum indegree. Every $x\in C$ is reached from $a$ in at most two steps
and reaches $b$ in at most two steps. Indeed, if $x$ beats $a$ and cannot
be reached from $a$ in two steps, $x$ beats all outneighbors of $a$ as
well, contradicting maximality. The reverse assertion follows by reversing
the edges. Any vertex on directed paths of the form $a\leadsto x\leadsto b$
belongs to $C$, because $b$ reaches $a$. Thus a circuit recognizing $C$
can hardwire $a,b$ and guess the two possible intermediate vertices.
Replacing the side membership test in the conditional anchor-choice
proof of Theorem~\ref{fc:efficient-quarter} with this circuit gives
$r(n-r)+O_L(n)$ bits for some
$r\le\lceil v\rceil+1$. The maximum of $r(n-r)$ in this range is at most
$u(n-u)+O(n)$. The circuit and its two hardwired vertices are used to
prove existence of a seed and index. Their descriptions are not needed
by the decoder.

For the lower bound let $N=n-1$. Given a cut on $\{0,1\}^N$, put
$p=|P|$ and assume $3\le p\le B-3$. Keep the $2^N$ old vertices with
prefix zero. Split the prefix-one vertices in lexicographic order
into sets $W,U,T$ of sizes $p,B-p,2^N-B$. The positive set is
$0P\cup U\cup T$ and the negative set is $0P^{\mathrm c}\cup W$.
Orient $U$ and $W$ by descending rank, reversing the extreme edge in
each. Each contains a directed Hamilton cycle since its size is at least
three. Give $T$ any public internal orientation. Put $T$ above every
other vertex, $U$ above all old vertices and $W$, and all old vertices
above $W$. Retain the old table among old vertices. Make just two
exceptions: for public representatives $u_0\in U,w_0\in W$ and
private representatives $c\in B^+_{\rm old},d\in B^-_{\rm old}$,
orient $0c\to u_0$ and $w_0\to0d$.

The exceptions preserve legality. Each old positive reaches $0c$ and
thus $U$, and $U$ can reach every old positive through a vertex other
than $u_0$. Each old negative can reach $W$ through a vertex other
than $w_0$, and $w_0$ reaches every old negative through $0d$.
Hence the two components adjacent to the padded cut are precisely
$U\cup0P$ and $0P^{\mathrm c}\cup W$. Their sizes are $B$ and
$2^N$. Membership on old vertices is unchanged.

For each allowed field let $G$ be the multilinear extension of the
public padding table, set to zero on old--old pairs. For fixed $n,B$
the data determining $G$ are $p,c,d$. With an old extension $F$,
simulate the new extension by
\[
 F'(b,x,b',y)=(1-b)(1-b')F(x,y)+G(b,x,b',y)+Z(b,x,b',y),
\]
where $Z$ is a uniformly random individual-degree-two polynomial
vanishing on the Boolean cube. Use one independent $Z$ per field and
retain these choices across the entire set of input runs. Conditional
on the old source and its extensions, $F'$ is uniform on the desired
fiber by translation, independently across fields. Fixing the masks
makes simulation deterministic and costs at most one old query per new
query. Unrestricted local computation permits the masks in the
description of the fixed-length simulator. For any all-core success
bound $\delta$ uniform over deterministic old strategies, the
unconditional probability of admissible padding \emph{and} all-core
success for a fixed padded strategy is at most $2^{3N+1}\delta$.
This follows by fixing the masks, summing over all triples $p,c,d$,
and then averaging over the masks.

Let $Q$ bound the query cost after $O(n)$ independent repetitions
using the same ordinary advice. Take
\[
 m=\min\{\lfloor\log_2 B\rfloor-2,\,
               \lfloor N/2\rfloor\},\quad K=2^m,\quad
 T_0=2^{N-m},\quad j=m-\lceil4\log_2(n+2)\rceil .
\]
If $j<1$, the claimed lower bound is nonpositive for a suitable
constant in the remainder. An instance with the required size exists
by taking two strong components of size $B$ and putting the remaining
vertices above them. Otherwise take the old hidden-core source with
$K$ buckets of size $T_0$. The fair core labels give $p\ge3$ with
probability $1-o(1)$, and $p\le K\le B/4$. Since $m\le N/2$ and $Q$
is polynomial, $T_0\ge32eQ$ for large $n$. Corollary~\ref{cor:saturation-good-rounds}
at discovery threshold $j$ gives the all-core estimate
\[
 \delta\le
 \left(\frac{32eQ}{3T_0}\right)^{K/4}
 +8\left(\frac{8eQ}{jT_0}\right)^j
 +\exp(-2^{-j-1}K/16)
 \le 2^{-m(N-m)+O_M(n\log n)}.
\]
Indeed $m-j=O(\log n)$, $\log Q=O_M(\log n)$, and $K2^{-j}$
is at least $(n+2)^4$.

If one advice word gives pointwise error $1/3$, its amplified decoder
is correct on all $K$ embedded core inputs with probability at least
$3/4$. Average the deterministic success bound over the fixed
amplification tapes and sum over the fewer than $2^{\ell+1}$ words of
length at most $\ell$. The probability of admissible padding with any
such successful word is at most $(4/3)2^{\ell+1+3N+1}\delta$, even when
the word is chosen after seeing the padded extensions. Taking
$\ell=m(N-m)-C_M n\log n$ makes this smaller than the probability of
admissible padding. This supplies the asserted cut and extension tuple.
Finally $|m-u|\le3$ throughout the nonvacuous range, so
$m(N-m)=u(n-u)+O_M(n\log n)$.
\end{proof}

\section{One oracle realizing the bounds}\label{fc:random-world-section}

\begin{theorem}[Relativized sharpness]\label{fc:random-world}
\label{thm:oracle}
There are a Boolean oracle $A$, a degree-two extension family
$\widetilde A$, and a $\mathrm P^A$-selective language $L$ with
$L\in\mathrm P^A/(n^2/4+O(n))$ such that the following hold for
almost every independent random Boolean oracle $R$. Write $O=A\oplus R$.
\begin{enumerate}
\item $\mathrm H^O$ holds. Every $\mathrm P^O$-selective language $L'$
has a deterministic polynomial-time decoder relative to $O$ with
$n^2/4+O_{L'}(n)$ ordinary advice bits.
\item There is a degree-two extension family $\widetilde O$ such that
$L\notin\mathrm{BPP}^{\widetilde O}/\lfloor cn^2\rfloor$ for each
$0<c<1/4$.
\end{enumerate}
In $\mathrm H^O$ the hard function belongs to $\mathrm E^O$ and the
competing circuits may use $O$ and $\mathrm{SAT}^O$ gates. The latter
decides satisfiability of an explicitly encoded Boolean oracle circuit.
Input wires and query lengths count toward circuit size.
\end{theorem}

\begin{proof}
\emph{The selector and its extensions.}
Use the active lengths $n_0=4$, $n_{i+1}=2^{n_i}$. At an active
length $n$, let $E$ and $P$ be the selected tournament and cut, put
$L_n=P$, and use two kinds of records:
\[
 A(00uv)=E(u,v)\quad(|u|=|v|=n),\qquad
 A(01x0^{2^n-n})=\mathbf1_P(x)\quad(|x|=n).
\]
Their lengths $2n+2$ and $2^n+2$ are disjoint throughout the tower.
Unspecified records are zero and inactive language slices are empty.
At an edge-record length extend over each field by
\begin{equation}\label{eq:tagproduct}
 \widetilde A(t,t',Y,Z)=(1-t)(1-t')\widetilde E(Y,Z).
\end{equation}
Use multilinear extensions at padded membership lengths and zero at
empty arities. At a skipped active length take an empty cut with a
fixed legal tournament. The selector compares two strings of equal
active length with one edge query. An active string is preferable
to an inactive one, and either of two inactive strings can be returned.
For different active lengths $r<s$ we have $2^r\le s$, so the
selector can read the shorter string's membership record and return
that string precisely when it is positive. Active lengths can be
recognized in polynomial time by iterating the tower up to the input
length. This gives $\mathrm P^A$-selectivity for every choice of cuts.

\emph{One base for almost every random oracle.}
Predetermine a schedule of clocked randomized classical interpreters $M$
with algebraic access, each repeated infinitely often. At a scheduled
active length take $m=\lfloor n/2\rfloor$, sample the Boolean hidden-core
source, and independently conditional on its table sample an extension
from each relevant degree-two fiber. One canonical encoding per field
suffices. Include all fields accessible to the machine within its
amplified clock, and complete the other fields multilinearly. The clock
allows the $O(n)$ repetitions used for ordinary advice. Choose the stage
length large enough that the current padded membership record and all
future records are out of reach. Sample the selected stages independently.

Fix any $R$ and condition on all earlier choices. For the tagged
union use at each arity
\[
 \widetilde O(t,y)=(1-t)\widetilde A(y)+t\,\mathrm{ML}(R)(y),
\]
where the last term uses the multilinear extension of the corresponding
$R$-slice. This is a valid degree-two extension. For the simulation an
algebraic evaluation, even at a non-Boolean tag, requires at most one
current extension evaluation by \eqref{eq:tagproduct}. The remaining
contributions are fixed before this stage. The finite bounds are uniform
in these auxiliary functions; their evaluation requires no queries to the
present hidden-core table.

Apply Corollary~\ref{cor:adaptive-advice} to $M$ at $m=\lfloor n/2\rfloor$,
with the auxiliary answers hardwired into the deterministic simulations,
and intersect its event with Lemma~\ref{lem:compressed-cover}. The joint
failure probability is $o(1)$, uniformly in the fixed auxiliary answers.
Advice may depend on the entire fixed oracle, as permitted there.

Choose the $i$th length in advance so that its failure probability
is at most $2^{-i}$. Borel--Cantelli gives both events on all
but finitely many selected stages, almost surely for this fixed $R$.
Fubini now fixes one base triple $(A,\widetilde A,L)$ for which this
holds for almost every $R$. The compressed cover gives polynomial-time
recognition using $m(n-m)+O(n)\le n^2/4+O(n)$ ordinary bits at the
selected lengths, with an empty-slice flag elsewhere. Finitely many
exceptions can be hardwired. At the infinitely many good stages of $M$
no word of length at most $n^2/4-C_Mn\log n$ gives pointwise error $1/3$.
For each $0<c<1/4$ this length exceeds $\lfloor cn^2\rfloor$ for large $n$,
which proves the second assertion.

\emph{Hardness and polynomial-time reconstruction in the same world.}
For each fixed base $A$, every size-$S$ circuit with $O$ and
$\mathrm{SAT}^O$ gates depends only on $A$ and on bits of $R$ at
lengths at most $(S+2)^b$ for an absolute $b$. This follows from
the explicit encoding of each queried circuit and its wires.
There are at most $2^{CS^2\log(S+2)}$ such circuit descriptions,
including hardwired constants. For $x\in\bits^n$ put
$f^O(x)=R(0\,0^{2^n}1x)$, which belongs to $\mathrm E^O$. Take
$S=\lfloor 2^{\eta n}\rfloor$ with $0<\eta<1/(4(b+1))$.
Conditional on the accessible short bits, the entire truth table
of $f^O_n$ is uniform. The probability that any size-$S$ circuit
computes it is at most $2^{CS^2\log(S+2)-2^n}$, summable in $n$.
Thus $\mathrm H^O$ holds almost surely.

The same oracle supplies the generator in
Theorem~\ref{fc:efficient-quarter} directly. For output length $N$
and fixed integer $k\ge1$, take $S=N^k$, $d=\lceil(2k+3)\log N\rceil$
and read the $N$ output bits of $G_{N,k}^R(z)$ at distinct $R$-addresses
encoding $(N,k,z,j)$, $z\in\bits^d$, $1\le j\le N$. Use a prefix $1$
and a self-delimiting encoding, padding its length beyond $(S+2)^b$
but polynomially in $N$ for each $k$. Conditional on all shorter
bits accessible to the circuits, the $2^d$ outputs are independent
and uniform. Hoeffding's inequality and the circuit count bound the
probability of discrepancy exceeding $1/4$ for any size-$S$ test by
$2^{CS^2\log(S+2)}\,2\exp(-2^d/8)$. This is summable in $N$.
Intersect over the countably many $k$. The generator is computable
in polynomial time with $O$. For a $\mathrm P^O$ selector the
weighted counting and block-index predicates of
Theorem~\ref{fc:efficient-quarter} are nondeterministic $O$-circuit
tests; their acceptance reduces to one $\mathrm{SAT}^O$ gate.
Applying the generator to the resulting tests gives the same
$n^2/4+O_{L'}(n)$ advice bound. Intersect these probability-one
properties of $R$ with the one already obtained for the fixed base.
\end{proof}

\appendix
\section{Equilibrium search bounds}\label{app:games}
\subsection{The lower bound with adaptive payoff queries}\label{app:probes}
A proposal algorithm may read $q$ payoff entries, adaptively, before
returning one profile. Write $P_\epsilon^{[q]}(N)$ for its optimal
worst-case success probability. No bound is imposed on computation
between queries. We prove the oblivious lower bound as the case $q=0$.

\begin{theorem}\label{g:probe}
Fix $0<\epsilon<1/2$. Set $n=\log_2N$, $c=1-H_2(1/2+\epsilon)$,
$B=2q+\lceil (n+2)^4/(\epsilon c)\rceil$ and
$s=\max\{0,\log_2(N/B)\}$. Then
\[
 -\log_2 P_\epsilon^{[q]}(N)
 \ge \frac{s^2}{4c}-O_\epsilon((s+1)\log(n+2)).
\]
In particular $q=N^{o(1)}$ preserves the leading exponent in
Theorem~\ref{g:main}. The lower bound holds for tournament games, whose
off-diagonal payoffs are zero or one and whose diagonal payoffs are $1/2$.
\end{theorem}

We give the dependence on the accuracy as well. For $0<\epsilon\le\epsilon_0<1/2$,
let $v=\log_2((n+2)/c)$. The proof gives the uniform error term
$O_{\epsilon_0}((s+1)v/c)$ in the displayed lower bound. In particular,
an implicit constant can be taken absolute for $\epsilon_0=1/4$.

\begin{lemma}[Weighted wins against one row]\label{g:tail}
Let $\beta=1/2+\epsilon$, $\eta=\epsilon/(n+2)$ and
$1\le h\le n/c$. If $w_1,\ldots,w_h$ are nonnegative and sum to one,
and $X_1,\ldots,X_h$ are independent fair bits, then for all sufficiently
large $n$, uniformly for $0<\epsilon\le\epsilon_0<1/2$,
\[
 \Pr\!\left[\sum_j w_jX_j>\beta+2\eta\right]
       \ge 2^{-ch-O_{\epsilon_0}(\log(h+2))}.
\]
\end{lemma}
\begin{proof}
Take $k=\lceil(\beta+2\eta)h\rceil+1$. For $h$ below a constant
depending only on $\epsilon_0$, use the event that all bits equal one.
For larger $h$, $k/h$ stays in a compact subinterval of $(0,1)$ and
$k\le h$. Conditional on exactly $k$ wins, the weighted sum has
mean $k/h\ge\beta+2\eta+1/h$ and lies in $[0,1]$. It exceeds
$\beta+2\eta$ with probability at least $1/h$. The probability of this
many wins is at least $2^{-h[1-H_2(k/h)]}/(h+1)$. Write
$k/h=\beta+\delta$ with $\delta=O(\eta+1/h)$. On the specified interval
the first derivative of $1-H_2$ at $\beta$ is $O_{\epsilon_0}(\epsilon)$
and its second derivative is bounded. Hence
\[
 h\bigl(1-H_2(k/h)-c\bigr)
      =O_{\epsilon_0}(h\epsilon\delta+h\delta^2)=O_{\epsilon_0}(1),
\]
using $h\le n/c$ and $c=\Theta_{\epsilon_0}(\epsilon^2)$. This proves
the bound for arbitrary weights.
\end{proof}

\begin{proof}[Proof of Theorem~\ref{g:probe}]
If $s<Cv$ the uniform bound follows from $P_\epsilon^{[q]}(N)\le1$
by enlarging its constant. Assume $s\ge Cv$, for a sufficiently large
$C$ depending only on $\epsilon_0$. Form $K=\lfloor\sqrt{N/B}\rfloor$
buckets of $d=\lfloor N/K\rfloor$ actions each. From each bucket choose
one action independently and uniformly. The resulting $K$ core actions
beat all noncore actions, including any leftover actions. Edges between
core actions are independent fair coins. Fix a public order on noncore
actions. Write $m=\log_2K$. Then $m=s/2+O(1)$, $d>B$ and
$\log_2((d-B)/B)=m+O(1)$.

Fix the algorithm's coins. When it queries an edge, reveal to it the
core status of both endpoints. After it outputs a pair, let $z$ be
the average of the two strategies, and expose for the analysis every
action with $z_i\ge b$, where $b=\eta c/(n+2)^3$. In total the set
$E$ of exposed names has size at most $B$. In each unresolved bucket
the hidden position is uniform on its untested names, independently
of the other unresolved buckets. Indeed, an inspected payoff can depend
on a random edge only after both core endpoints have been identified.
For $h$ the number of exposed core actions, a union bound over exposure
times gives, for every integer $h_0\ge1$,
\[
 \Pr[h\ge h_0]\le \binom B{h_0}(d-B)^{-h_0}.
\]
Condition on the enlarged transcript. The total probability mass $Y$ that
$z$ assigns to unexposed core actions is a sum of independent terms
in $[0,b]$, with mean at most $1/(d-B)$. The chord bound
$\exp(w)\le 1+(e-1)w$ for $0\le w\le1$ yields
\[
 \Pr[Y>\eta]\le
 \exp\!\left(-\eta/b+\frac{e-1}{b(d-B)}\right)
 \le \exp[-\Omega((n+2)^3/c)].
\]
Here $1/(d-B)=o(\eta)$ in the range under consideration since
$B\ge1/b$ and $(d-B)/B=\Theta(K)$.

Fix the remaining core locations. Suppose $Y\le\eta$ and
$h<h_0=\lfloor(m-C'v)/c\rfloor$, with $C'$ sufficiently large and
$C$ chosen to make $h_0\ge1$. No unexposed core action was queried.
Its edges to the $h$ exposed core actions are fresh fair bits, and the
collections of these edges are independent for different challengers.
If the exposed core mass $a$ of $z$ is positive, normalize it to a
distribution on those $h$ actions. By Lemma~\ref{g:tail}, each
challenger beats more than a $\beta+2\eta$ fraction of the normalized
mass with probability at least $2^{-ch-O_{\epsilon_0}(v)}$. Its payoff
against $z$ then exceeds
\[
          1-a-\eta+a(\beta+2\eta)\ \ge\ \beta
\]
for large $n$. If $a=0$, every challenger already earns at least
$1-\eta>\beta$. There are $K-h\ge K/2$ challengers. By the choice of
$C'$, the probability that none earns more than $\beta$ is at most
$\exp[-\Omega((n+2)^3/c)]$. Thus the success probability is at most
\[
  2\exp[-\Omega((n+2)^3/c)]
       +\left(\frac{B}{d-B}\right)^{h_0}
 \le 2^{-s^2/(4c)+O_{\epsilon_0}((s+1)v/c)}.
\]
The same distribution of games works for every fixing of the coins.
Averaging gives the theorem and the lower bound in Theorem~\ref{g:main}.
\end{proof}

\paragraph{Accuracy and advice.}
For $q=0$, $s=n-O_{\epsilon_0}(v)$. Lemma~\ref{g:count} gives an
upper cost $n^2/(4c)+O(n/c+\log(n/c))$. To obtain a universal list for
shrinking $\epsilon$, round payoffs using an even grid denominator
$D=\Theta(n/\epsilon)$ and tolerance $\epsilon-1/D$ in the sampling
argument. For $\epsilon\le1/4$ this changes $c$ by a relative $O(1/n)$,
because $c=\Theta(\epsilon^2)$ and $c'(\epsilon)=O(\epsilon)$.
Consequently for $\log(1/\epsilon)=o(n)$ both logarithmic costs equal
\[
       \frac{n^2}{4c}+O\!\left(\frac n c\log\frac{n+2}{c}\right)
       \qquad(0<\epsilon\le1/4).
\]
The same leading term for one proposal persists when
$\log(q+1)+\log(1/\epsilon)=o(n)$. The lower bound further allows
ordinary game-dependent advice: fix each advice word and sum the
success probabilities on the common hard distribution. Constant success
with $\ell$ bits requires $\ell\ge s^2/(4c)-O_{\epsilon_0}((s+1)v/c)$.
An index in the universal list gives the matching leading advice upper
bound when $q=N^{o(1)}$ and $\epsilon$ is fixed.

\subsection{Security strategies in general zero-sum games}\label{g:general}
Consider the
same one-player task for an arbitrary $N$-by-$N$ payoff matrix
$A\in[0,1]^{N\times N}$ of value $\min_p\max_i(Ap)_i=1/2$. The column
player seeks a security strategy $p$ with $\max_i(Ap)_i\le\beta=1/2+\epsilon$.
Write $P^{\rm sec}_\epsilon(N)$ for the optimal worst-case success probability
of a game-independent proposal and $L^{\rm sec}_\epsilon(N)$ for the minimum
size of a universal list of proposals.

\begin{theorem}\label{g:general-sec}
For fixed $0<\epsilon<1/2$, $n=\log N$ and $c=1-H_2(\beta)$,
\[
 \frac{n^2}{c}-O_\epsilon(n\log n)
 \ \le\ -\log P^{\rm sec}_\epsilon(N)
 \ \le\ \log L^{\rm sec}_\epsilon(N)
 \ \le\ \frac{n^2-n\log n}{c}+O_\epsilon(n).
\]
The lower bound holds for binary matrices of value exactly $1/2$.
\end{theorem}
\begin{proof}
For the upper bound sample $r=\lceil(n+1)/c\rceil$ times from an optimal
column strategy. The bounded-variable Chernoff bound gives probability at most
$2^{-cr}$ that a fixed row exceeds $\beta$. Thus the list of the
$\binom{N+r-1}{r}$ empirical strategies contains a successful one.
Its logarithm is $rn-r\log r+O(r)$ by Stirling's formula.
Drawing uniformly from a
universal list succeeds with probability at least the reciprocal of its size.

For the lower bound choose a uniform set $C$ of $K=2\lceil n^8/2\rceil$
columns, for $N$ sufficiently large. Choose $\lfloor N/2\rfloor$
independent uniform $K/2$-subsets of $C$. Each supplies two rows, its
indicator on $C$ and the complementary indicator, both equal to one outside
$C$. Duplicate the first row if $N$ is odd. The uniform strategy on $C$
pays exactly $1/2$ on every row. A uniformly random member of any row
pair earns at least $1/2$ on every column. Hence the value is $1/2$.

Fix any proposed distribution $p$. Its heavy set
$H_0=\{j:p_j\ge 1/(Kn^2)\}$ has size at most $Kn^2$.
Write $H=H_0\cap C$, $h=|H|$. The planted mass outside $H$ is at most
$n^{-2}$. Conditional on $C$, let $Z$ be the payoff of the first row in
one pair against $p$. Given that the row contains exactly $k$ ones on $H$,
exchangeability gives
\[
 \mathbb E[Z\mid k,C]\ge (k/h)p(H)+p(C^c)\ge k/h-n^{-2}
 \qquad(h>0).
\]
For $h=O_\epsilon(n)$ large enough, take $k=\lceil\beta h\rceil+1\le h$.
Since $0\le Z\le1$, this gives
$\Pr[Z>\beta\mid k,C]=\Omega_\epsilon(1/h)$.
The probability of this hypergeometric layer is at least
$\binom hk2^{-h}\exp(-O(h^2/K))$: the probability of any fixed pattern is
$(K/2)_k(K/2)_{h-k}/(K)_h$, using falling factorials.
The binomial entropy estimate gives
$\Pr[Z>\beta\mid C]\ge 2^{-ch-O_\epsilon(\log(h+2))}$.
For bounded positive $h$ use the all-ones pattern, and for $h=0$ the
outside mass exceeds $\beta$.

Choose $h_0=\lfloor(n-B_\epsilon\log n)/c\rfloor$ with $B_\epsilon$
sufficiently large. When $h<h_0$ the $\lfloor N/2\rfloor$ independent
first rows reject $p$ except with probability $\exp(-\Omega(n^3))$.
Otherwise a union bound gives
\[
 \Pr_C[h\ge h_0]\le \binom{\lfloor Kn^2\rfloor}{h_0}(K/N)^{h_0}
 \le 2^{-n^2/c+O_\epsilon(n\log n)} .
\]
Averaging over arbitrary game-independent proposals proves the lower bound.
\end{proof}

\subsection{Symmetric games with arbitrary payoffs}\label{g:symmetric-general}
For a symmetric bimatrix game let $A\in[0,1]^{N\times N}$ be arbitrary.
The two payoffs at $(i,j)$ are $A_{ij}$ and $A_{ji}$. Write
$P_\epsilon^{\rm sym}(N)$ and $L_\epsilon^{\rm sym}(N)$ for the
proposal probability and list size in Theorem~\ref{g:main} with this
larger class of games. Both players may again use arbitrary mixed
strategies in a proposal.

\begin{theorem}\label{g:gensym}
For every fixed $0<\epsilon<1/2$, with $n=\log N$,
\[
 \frac{1}{4[1-H_2(1/2+\epsilon)]}
 \ \le\ \liminf_{N\to\infty}\frac{-\log P_\epsilon^{\rm sym}(N)}{n^2}
 \ \le\ \limsup_{N\to\infty}\frac{\log L_\epsilon^{\rm sym}(N)}{n^2}
 \ \le\ \frac{\ln 2}{8\epsilon^2}.
\]
If $\epsilon=\epsilon(N)\to0$ and $\log(1/\epsilon)=o(n)$, both
quantities have the same leading term:
\[
 -\log P_\epsilon^{\rm sym}(N)
 =\left(\frac{\ln2}{8}+o(1)\right)\frac{n^2}{\epsilon^2},
 \qquad
 \log L_\epsilon^{\rm sym}(N)
 =\left(\frac{\ln2}{8}+o(1)\right)\frac{n^2}{\epsilon^2}.
\]
\end{theorem}
\begin{proof}
The lower bounds follow by restricting the games to the constant-sum
subclass. We prove the upper bound directly. For a common mixed action
$p$ let $u(p)=p^{\mathsf T}Ap$ be the payoff at $(p,p)$. A symmetric
proposal $(z,z)$ has regret $\max_i (Az)_i-u(z)$ for either player.
Unlike the constant-sum case, its baseline $u(z)$ varies with the sample.

Take $n$ large and put
\[
 \Delta=\epsilon n^{-1/4},\qquad e=\epsilon-\Delta,\qquad
 t=4e,\qquad d=2e^2/\ln2.
\]
Use the continuous functions
\[
 q_p(i)=\exp[-t(u(p)+e)]\bigl(1+(\exp t-1)(Ap)_i\bigr).
\]
If $X$ is Bernoulli with mean $u(p)$, then
$\sum_i p_iq_p(i)=e^{-te}\mathbb E\exp(t(X-\mathbb EX))$.
Hoeffding's lemma gives
\[
 \sum_i p_iq_p(i)\le \exp(-te+t^2/8)=2^{-d}.
\]
Lemma~\ref{fp:scores} applies with $d$ in
place of $c$. Indeed, the first coordinate at a fixed point gives
$N^{-1/J}\le Z^{-1/J}\sum_i p_iq_p(i)\le Z^{-1/J}2^{-d}$. Thus the
same endpoint argument yields $Z=1/2$, positive $J\le(n+1)/d$, and
\[
 p_i=q_p(i)^J/Z,\qquad q_p(i)<1,\qquad
 \sum_i p_i^{1+1/J}\le 2^{1/J-d}.
\]

Put $r_0=\lceil32/\Delta^2\rceil$ and
$r=\max\{\lceil J\rceil,r_0\}$. Draw $r$ actions independently from
$p$ and let $z$ be their empirical distribution.
The exponential Markov argument of Lemma~\ref{g:count} now bounds
$\Pr[\max_i(Az)_i>u(p)+e]$ by $\sum_i q_p(i)^r\le1/2$.
It remains to bound the common payoff $u(z)$. Expanding its double
sum over the sampled actions gives
\[
 \mathbb E u(z)=(1-1/r)u(p)+(1/r)\sum_i p_iA_{ii}.
\]
Its bias is at most $1/r$. Changing one sampled action changes $u(z)$
by at most $2/r$, since only terms involving its index can change.
As $1/r\le\Delta/2$, the bounded-difference inequality gives
\[
 \Pr[u(z)<u(p)-\Delta]\le\exp(-r\Delta^2/8)<1/4 .
\]
Thus at least a quarter of the product measure consists of successful
tuples.

When $r=\lceil J\rceil$, H\"older's inequality as in Lemma~\ref{g:count}
gives at least $2^{dr^2-O(r+1)}$ successful tuples, with an absolute
implicit constant for $0<\epsilon\le1/2$. Their ambient density is at
least $2^{-n^2/(4d)-O(n/d)}$. If $r=r_0>\lceil J\rceil$, the existence
of even one successful tuple gives ambient density at least $2^{-nr_0}$.
Here $nr_0=O(n^{3/2}/\epsilon^2)=o(n^2/\epsilon^2)$, uniformly in
$\epsilon$. An oblivious sampler chooses the length uniformly through
$\max\{\lceil(n+1)/d\rceil,r_0\}$ and samples a uniform ambient tuple.
The cost of guessing the length in the exponent is
$O(\log(n+2)+\log(1/\epsilon))$. Since
$d=(2\epsilon^2/\ln2)(1-n^{-1/4})^2$, these bounds give the claimed
upper coefficient.

For a universal list, round entries of $A$ to a grid of denominator
$D=\lceil n/\epsilon\rceil$. Each player's expected payoff and its
best-deviation payoff each change by at most $1/D$. Apply the sampler to
the rounded games with tolerance $\epsilon-2/D=\epsilon(1-O(1/n))$.
There are at most $(D+1)^{N^2}$ such games. As in Theorem~\ref{g:main},
sampling a finite list adds at most $2n+O(\log\log(D+1))$ bits to the
logarithm of its size. This preserves the stated leading terms.
\end{proof}

\subsection{General bimatrix games}\label{g:two}
Let $P_\epsilon^{\rm bi}(N)$ and $L_\epsilon^{\rm bi}(N)$ denote the
oblivious success probability and universal list size for ordinary
additive $\epsilon$-Nash equilibria of games with two arbitrary payoff
matrices $A,B\in[0,1]^{N\times N}$, in row and column order.
\begin{theorem}\label{g:bimatrix-two}
Write $n=\log N$. For every fixed $0<\epsilon<1/2$, and also for
$\epsilon\to0$ with $\log(1/\epsilon)=o(n)$,
\[
 -\log P_\epsilon^{\rm bi}(N)\ \le\ \log L_\epsilon^{\rm bi}(N)
 \ \le\ \left(\frac{\ln 2}{2}+o(1)\right)\frac{n^2}{\epsilon^2}.
\]
\end{theorem}
\begin{proof}
Put $\Delta=\epsilon n^{-1/4}$, $e=\epsilon-\Delta$, $t=4e$,
$d=2e^2/\ln2$, and $\zeta=1/8$. For mixed strategies $p,q$ write
$a=p^{\mathsf T}Aq$, $b=p^{\mathsf T}Bq$ and define
\[
 g_i=\sum_jq_j\exp\bigl(t(A_{ij}-a-e)\bigr),\qquad
 h_j=\sum_ip_i\exp\bigl(t(B_{ij}-b-e)\bigr).
\]
Under $p\otimes q$ each payoff has range in $[0,1]$ and mean equal
to its baseline. Hoeffding's lemma gives
$\sum_i p_i g_i,\ \sum_j q_j h_j\le\exp(-te+t^2/8)=2^{-d}$.

There are $p,q$ and $J,L>0$ satisfying
\[
 Z=\sum_i g_i^J=\zeta,\quad W=\sum_j h_j^L=\zeta,\qquad
 p_i=g_i^J/Z,\quad q_j=h_j^L/W .
\]
Indeed put $U=(n+4)/d$ and apply Brouwer on
$\Delta_N^2\times[0,U]^2$. Normalize the two power vectors for the
first two coordinates, and project $J+\log(Z/\zeta)$ and
$L+\log(W/\zeta)$ onto $[0,U]$ for the last two.
All entries are positive and continuous. At exponent zero its sum
is $N$, so its coordinate increases. At a fixed point with $J>0$,
\[
 2^{-n/J}\le\sum_i p_i^{1+1/J}
             = Z^{-1/J}\sum_i p_i g_i \le Z^{-1/J}2^{-d},
\]
which gives $Z\le2^{n-dJ}$. At $J=U$ this is less than $\zeta$,
so that coordinate decreases. The same inequalities hold for $L,W$.
Both exponents are therefore interior, and their fixed-point equations
give the claimed normalizations. In particular
$J,L\le(n+3)/d$ and $g_i,h_j<1$.

Put $m_0=\lceil 16/\Delta^2\rceil$.
Draw $s=\max\{\lceil L\rceil,m_0\}$ row actions from $p$ and
$r=\max\{\lceil J\rceil,m_0\}$ column actions from $q$, all independently.
For the empirical profile $(x,y)$, exponential Markov and a union bound give
\[
 \Pr[\max_i(Ay)_i>a+e]\le\sum_i g_i^r\le\zeta,\qquad
 \Pr[\max_j(x^{\mathsf T}B)_j>b+e]\le\sum_j h_j^s\le\zeta .
\]
The payoffs $x^{\mathsf T}Ay,x^{\mathsf T}By$ have means $a,b$.
Changing a sampled row or column action changes either payoff by at
most $1/s$ or $1/r$, respectively. The bounded-difference inequality
therefore bounds each lower tail at distance $\Delta$ by
$\exp(-2\Delta^2/(1/s+1/r))\le e^{-16}$. With probability at least
$1/2$ the empirical pair is an $\epsilon$-Nash equilibrium.

Row tuples whose conditional success probability is at least $1/4$
have $p^{\otimes s}$-mass at least $1/3$. H\"older and the displayed
power-moment bound give at least
\[
 (1/3)^{J+1}\left(\sum_i p_i^{1+1/J}\right)^{-sJ}
 \ \ge\ (1/3)^{J+1}\zeta^s 2^{dJs}
\]
such rows. For each of them the same argument gives at least
$(1/4)^{L+1}\zeta^r2^{dLr}$ successful column tuples. Thus the
negative logarithm of the density among ambient ordered tuples is at most
\[
 n(s+r)-d(Js+Lr)+O(J+L+s+r+1)
 \ \le\ n(J+L)-2dJL+O(nm_0+n/d).
\]
The cross term reflects the two roles of $J$: it supplies the entropy
bound on row actions and the sample length for column actions. The
bilinear expression is largest at a corner of
$[0,(n+3)/d]^2$, where it is at most $n(n+3)/d$.
Guessing both lengths uniformly up to
$\max\{\lceil(n+3)/d\rceil,m_0\}$ costs
$O(\log(n+2)+\log(1/\epsilon))$ additional bits in the exponent.
Since $nm_0=O(n^{3/2}/\epsilon^2)$, this proves the probability bound
with the stated coefficient.

For the list, discretize both payoff matrices on a grid with
denominator $D=\lceil n/\epsilon\rceil$. Each deviation gain changes
by at most $2/D$. Use the sampler at tolerance $\epsilon-2/D$ and
take independent proposals to hit every rounded game. There are at
most $(D+1)^{2N^2}$ games on the grid. The logarithm of the required
list size exceeds the negative logarithm of single-proposal success
by at most $2n+O(\log\log(D+1)+1)$, which preserves the leading term.
\end{proof}

\begin{remark}[The bimatrix gap]\label{g:bimatrix-gap}
Theorem~\ref{g:bimatrix-two} has no matching lower bound. Zero-sum games
give one: if $B=\mathbf1\mathbf1^{\mathsf T}-A$ and $A$ has value $1/2$, an
$\epsilon$-Nash pair $(x,y)$ satisfies $\max_i(Ay)_i\le u+\epsilon$ and
$\min_j(x^{\mathsf T}A)_j\ge u-\epsilon$, so $u\le1/2+\epsilon$ and $y$ is
$2\epsilon$-secure. Theorem~\ref{g:general-sec} at accuracy $2\epsilon$
therefore gives $-\log P^{\rm bi}_\epsilon(N)\ge n^2/c_{2\epsilon}
-O_\epsilon(n\log n)$, which is $(\ln2/8+o(1))\,n^2/\epsilon^2$ for small
$\epsilon$. A factor four remains between this and the upper bound, and we
do not know the sharp constant for general bimatrix games.
\end{remark}

\begin{remark}[The Daskalakis--Papadimitriou packing for ordinary regret]
\label{g:dp09-packing}
In the hard family of~\cite[Section~3]{DP09}, each game hides an even-sized
set $S$ of $\ell$ rows among $N=\binom{\ell}{\ell/2}$ rows. The columns
are indexed by the half-size subsets $T$ of $S$. Change the row payoff
outside $S$ from $-1$ to $0$. The resulting payoffs are
$(\mathbf 1_{i\in T},1-\mathbf 1_{i\in T})$ at every entry $(i,T)$.
The uniform row strategy $u_S$ earns $1/2$ against every column.
At an ordinary $\epsilon$-Nash profile with row marginal $x$, the row
player therefore earns at least $1/2-\epsilon$. The column player's
regret bound then gives $x(T)\ge 1/2-2\epsilon$ for every $T$.

Let $D=\sum_{i\in S}(1/\ell-x_i)_+=\|x-u_S\|_1/2$.
The largest $\ell/2$ numbers $1/\ell-x_i$ sum to at least $D/2$.
Indeed, if at least half are positive, the largest half captures at least
half their sum. Otherwise, filling the half-set with the least negative
entries uses at most half the total negative mass, which is at most $D$
since $\sum_{i\in S}(1/\ell-x_i)=1-x(S)\ge0$.
For this half-set $T$ we have $D/2\le 1/2-x(T)\le 2\epsilon$.
Thus $\|x-u_S\|_1\le8\epsilon$. Claim~3 of~\cite{DP09} packs
$\Omega(2^{(0.8-34\epsilon)n^2})$ disjoint balls of this radius.
A proposal can succeed for at most one packed game, giving the stated
lower bound for $0<\epsilon<2/85$.
\end{remark}

\section*{AI disclosure}
Claude assisted in developing the fully adaptive degree-two discovery
argument,  the derandomized anchor selection
under $\mathrm H$,  and the consequences
for tournaments  in Section~2.1. ChatGPT (Codex)
assisted in developing and checking the the blockwise
derandomization, the algebraic lower bound and the oracle construction.
Both assisted in drafting and revising the exposition.
\end{document}